\documentclass[11pt]{article}
\usepackage[a4paper,margin=1in]{geometry}
\usepackage{amsmath,amssymb,amsthm,mathtools,bm}
\usepackage{booktabs}
\usepackage{enumitem}
\usepackage{algorithm}
\usepackage[noend]{algpseudocode}
\usepackage{xcolor}
\usepackage{graphicx}
\usepackage[colorlinks=true,linkcolor=blue,citecolor=blue,urlcolor=blue]{hyperref}

\newtheorem{proposition}{Proposition}[section]
\newtheorem{lemma}[proposition]{Lemma}
\newtheorem{theorem}[proposition]{Theorem}
\newtheorem{corollary}[proposition]{Corollary}
\newtheorem{remark}[proposition]{Remark}
\newlist{hypAitems}{enumerate}{1}
\setlist[hypAitems,1]{
  label=\textbf{(A\arabic*)},
  ref=\arabic*,
  leftmargin=3.2em,
  itemsep=0.35em
}
\newlist{hypHitems}{enumerate}{1}
\setlist[hypHitems,1]{
  label=\textbf{(H\arabic*)},
  ref=\arabic*,
  leftmargin=3.2em,
  itemsep=0.35em
}
\newcommand{\hypH}[1]{\textup{(H\ref{#1})}}
\newcommand{\hypL}[1]{\textup{(L\ref{#1})}}
\newlist{hypLitems}{enumerate}{1}
\setlist[hypLitems,1]{
  label=\textbf{(L\arabic*)},
  ref=\arabic*,
  leftmargin=3.2em,
  itemsep=0.35em
}
\newcommand{\hypA}[1]{\textup{(A\ref{#1})}}
\newcommand{\R}{\mathbb{R}}
\newcommand{\E}{\mathbb{E}}
\newcommand{\V}{\mathbb{V}}
\newcommand{\cP}{\mathcal{P}}
\newcommand{\cO}{\mathcal{O}}
\newcommand{\KL}{\operatorname{KL}}
\newcommand{\Wtwo}{\mathcal{W}_2}
\newcommand{\dd}{\mathrm{d}}
\newcommand{\eps}{\varepsilon}
\newcommand{\filter}{\eta}
\newcommand{\predictor}{\rho}

\newcommand{\law}{\operatorname{Law}}

\begin{document}
\begin{center}
{\Large \textbf{On Sequential Entropic Optimal Transport Filter and its Multilevel Variants}}

\vspace{0.5cm}

KE ZHAO$^{1}$\& AJAY JASRA$^{2}$ 

{\footnotesize $^{1}$School of Data Science,  The Chinese University of Hong Kong, Shenzhen,  Shenzhen,  CN.}\\
{\footnotesize $^{2}$Computing \& Mathematical Sciences Division,  Mohamed Bin Zayed University of Artificial Intelligence,  Abu Dhabi, UAE.}\\
{\footnotesize E-Mail:\,}
\texttt{\emph{\footnotesize kezhao@link.cuhk.edu.cn}};
\texttt{\emph{\footnotesize Ajay.Jasra@mbzuai.ac.ae}};

\end{center}

\begin{abstract}
We consider the problem of filtering partially observed diffusions associated to
discrete time observations. This problem requires the development of advanced
numerical methods such as particle filters.
In this article we develop a new sequential entropic optimal transport filter (EOTF).
The main idea is to sample from successive couplings of the filters.
To achieve this,
we adopt the approach of an 
entropic optimal transport (EOT) update implemented through a learned Schr\"odinger
potential, in the spirit of the light Schr\"odinger bridge methodology
\cite{korotin2024light}. 
Based on this, we develop a multilevel version of this filter (ML-EOTF)
in which consecutive discretization levels are coupled at the terminal time
by a scaled EOT plan with level-dependent regularization
$\eps_l=\eps_0 h_l^\kappa$ with discretization stepsize $h_l$. Under mild conditions, we show that the level-difference estimator with exact
Schr\"odinger potentials has variance of order
$N_l^{-1}\{h_l^{\beta}+\eps_l\}$, where $\beta$ is the strong-error for the discretized SDE. The analysis is then extended to learned potentials, with all
approximation and optimization errors summarized by a joint
Kullback--Leibler discrepancy whose contribution to the mean squared error
is explicitly quantified. Combined with a fixed-time weak-error estimate, this
yields the canonical multilevel complexity $\cO(\delta^{-2}|\log\delta|^2)$ to achieve an MSE of $\delta^2$
for Euler--Maruyama discretizations.  Numerical experiments validate the results.
\end{abstract}



\section{Introduction}
\label{sec:intro}

State-space models provide a standard probabilistic framework for
sequential data arising in signal processing, engineering and finance
\cite{cappe2005inference,doucet2001sequential}.
For nonlinear and non-Gaussian models, particle filters (PFs) are among
the most widely used methods for approximating the filtering recursion
\cite{gordon1993novel,delmoral2004feynman}.
In this article, we consider the partially observed diffusion setting,
in which the latent signal is a continuous-time diffusion observed only
at discrete times. After time discretization, the model becomes a
standard discrete-time state-space model to which particle filtering
can be applied; see for instance \cite{jasra2017mlpf,jasra2018mlpf,ub_filt}.
Controlling the discretization bias, however, may require a fine time
grid, so repeated propagation of the particle system can constitute a
substantial computational cost. To reduce it, one can use the idea of
multilevel Monte Carlo (MLMC) by combining a cheap
coarse approximation with increasingly fine correction terms
\cite{giles2008mlmc,giles2015mlmc}. This idea has been adapted to
sequential filtering through the multilevel particle filter (MLPF)
\cite{jasra2017mlpf}, where the key issue is to couple the fine and
coarse particle systems with a delicate design of resampling so that the variance of each level difference
decays sufficiently rapidly. 

An alternative to  resampling in PF is to transport the propagated
ensemble to the likelihood-weighted target ensemble, see \cite{reich2013nonparametric,gregory2016metpf,gregory2017seamless,sen2018coupling,aljarrah2023optimal,corenflos2021differentiable} and references therein. Pioneered by \cite{reich2013nonparametric}, a discrete
optimal transport (OT) solver is used though the cost is rather high, with $\mathcal{O}(N^3\log N)$ for $N$ particles; see also \cite{bertsimas1997introduction}. An entropic
regularization in OT can lead to \(\mathcal{O}(N^2)\) cost using Sinkhorn iterations 
\cite{cuturi2013sinkhorn,peyre2019computational}, which is used in the differential particle filter (DPF) in \cite{corenflos2021differentiable}. The primary focus of \cite{corenflos2021differentiable} is the differentiability of the filter with which one can combine neural network training in a more complicated setting. By contrast, the focus of the present article is to
use a learned EOT conditional kernel as the filtering update with lower cost and to
analyze its multilevel extension as follows.

EOT is intimately connected with the Schr\"odinger bridge problem. For
quadratic costs, the entropic plan admits a Schr\"odinger-potential
representation and can equivalently be interpreted through a mixture of
bridges; see, for example,
\cite{leonard2014survey,ghosal2022stability,nutz2022entropic,korotin2024light,gushchin2024matching} and also some recent analysis \cite{delmoral2026sinkhorn,delmoral2026bridging}. From a
computational viewpoint, the light Schr\"odinger bridge methodology
\cite{korotin2024light,gushchin2024matching} parameterizes the potential
by a Gaussian mixture, yielding a tractable conditional transport kernel
with near $\mathcal{O}(N)$ cost.
Based on the  representation of \cite{korotin2024light}, we introduce a sequential entropic
optimal transport filter (EOTF). Using the same idea, we transport the  particles in the coarse level to the fine level in the terminal coupling, which is rather novel in multilevel methods \cite{giles2015mlmc,jasra2017mlpf,gregory2016metpf,gregory2017seamless} since we do not necessarily ask to couple them at every stage. The coupling idea is not intrinsically restricted
to diffusion models. It can be used whenever fine- and coarse-level
particles are available for two adjacent target distributions.

The contributions of this article are threefold. First, we propose the
sequential entropic optimal transport filter (EOTF), which replaces the
resampling step of a conventional particle filter by sampling from a
learned EOT conditional kernel. 
Second, we develop the multilevel EOTF (ML-EOTF), which couples terminal
fine- and coarse-level ensembles through a scaled EOT plan with
level-dependent regularization \(\eps_l=\eps_0h_l^\kappa\). This scaling
allows the entropic displacement to be balanced against the
discretization error. Third, we provide a rigorous theoretical analysis
of the proposed methods, with main results in Theorem~\ref{thm:complexity}, saying that the multilevel complexity is $\cO(\delta^{-2}|\log\delta|^2)$ if one wants an MSE of $\delta^2$
for Euler--Maruyama discretizations. The theoretical predictions are
supported by experiments on synthetic and real data.

The remainder of the paper is organized as follows. Section~\ref{sec:setup} introduces
the filtering model and the EOT formulation that we will use. Section~\ref{sec:method} presents some basic ideas of PF and
MLPF; and propose our new methodology including EOTF and ML-EOTF algorithms. Section~\ref{sec:theory} develops the main theory, while Section~\ref{sec:numerics}
reports numerical experiments. Technical proofs and additional
computational details are deferred to the Appendix~\ref{app:proofs} and \ref{app:implement}.

\section{Problem setup and preliminaries}\label{sec:setup}

\subsection{Partially observed diffusion}

Let $(\Omega,\mathcal F,\mathbb P)$ support a $d_w$-dimensional Brownian
motion $(W_t)_{t\geq0}$.  The latent process $(X_t)_{t\geq0}$ takes values in
$\R^d$ and solves
\[
\dd X_t=a(X_t)\,\dd t+b(X_t)\,\dd W_t,
 \qquad X_0\sim\mu_0,
\]
where $a:\R^d\to\R^d$ and $b:\R^d\to\R^{d\times d_w}$.  Observations are
available at integer times. Conditionally on $X_p=x_p$, the observation
$Y_p=y_p$ has likelihood
\[
 G_p(x_p):=g(y_p\mid x_p),\qquad p\geq1.
\]
The observations $y_{1:n}$ are fixed throughout.

Write $M^\infty(x,\dd x')$ for the exact transition kernel over one
observation interval. At level $l\geq0$, it is replaced by a time-discretized
kernel $M^l(x,\dd x')$ with step size
\[
h_l=2^{-l}.
\]
For Euler--Maruyama, one observation interval contains $h_l^{-1}$ steps,
\[
X_{m+1}^l=X_m^l+a(X_m^l)h_l+b(X_m^l)\Delta W_m^l,
 \qquad \Delta W_m^l\sim\mathcal N(0,h_lI_{d_w}).
\]
In multilevel schemes \cite{giles2015mlmc,giles2008mlmc}, we usually use the aggregation of fine-level Brownian increments in coarse level
\begin{equation}
 \Delta W_k^{l-1}=\Delta W_{2k}^l+\Delta W_{2k+1}^l.
 \label{eq:brownian-coupling}
\end{equation}

The level-$l$ joint law of $X_{1:n}^l$ and $y_{1:n}$ is proportional to
\[
\mu_0(\dd x_0)\prod_{p=1}^n
 M^l(x_{p-1},\dd x_p)G_p(x_p).
\]
For a test function $\varphi:\R^d\to\R$, the filtering objective is
\[
\filter_n^l(\varphi)
 :=\E[\varphi(X_n^l)\mid y_{1:n}]
 =\int\varphi(x)\filter_n^l(\dd x).
\]
The predictor and filter satisfy
\begin{align*}
\predictor_n^l(\dd x)
 &=\int M^l(x',\dd x)\filter_{n-1}^l(\dd x'),\\
 \filter_n^l(\dd x)
 &=\frac{G_n(x)\predictor_n^l(\dd x)}
 {\predictor_n^l(G_n)}.
\end{align*}

For a probability measure \(\mu\) satisfying \(\mu(G_p)>0\), define the
Bayesian update map
\begin{equation}
 \Psi_p(\mu)(\dd x)
 :=\frac{G_p(x)\mu(\dd x)}{\mu(G_p)}.
 \label{eq:bayes-map}
\end{equation}
Then \(\filter_p^l=\Psi_p(\predictor_p^l)\). For a transition kernel \(M\)
and a measurable test function \(\psi\), write
\[
 M\psi(x):=\int_{\R^d}\psi(x')M(x,\dd x'),
 \qquad
 \mu M(\dd x'):=\int_{\R^d}M(x,\dd x')\,\mu(\dd x),
\]
the latter being the action of \(M\) on a measure \(\mu\).
Whenever \(M^l(x,\cdot)\) is absolutely continuous with respect to
Lebesgue measure, its density is denoted by \(m^l(x,\cdot)\), namely
\[
 M^l(x,\dd y)=m^l(x,y)\,\dd y.
\]
More generally, if a probability measure \(\mu\) has a Lebesgue density,
we write \(\mu(x)\) for its Radon--Nikodym derivative, so that
\(\mu(\dd x)=\mu(x)\,\dd x\). This convention is used only when existence
of the density has been established.
The multilevel identity is
\[
\filter_n^L(\varphi)
 =\filter_n^0(\varphi)+\sum_{l=1}^L
 \{\filter_n^l(\varphi)-\filter_n^{l-1}(\varphi)\}.
\]

\subsection{Entropic optimal transport}

Let $\mu,\nu\in\cP_2(\R^d)$ and let $\Pi(\mu,\nu)$ denote their couplings.
For $c(x,y)=\frac12\lVert x-y\rVert^2$, the EOT problem is
\begin{equation}
 \pi^\eps\in\arg\min_{\pi\in\Pi(\mu,\nu)}
 \left\{\int c(x,y)\pi(\dd x,\dd y)
 +\eps\KL(\pi\Vert\mu\otimes\nu)\right\}.
 \label{eq:eot-primal}
\end{equation}
Up to factors depending only on $x$ or $y$, which can be absorbed into the
Schr\"odinger potentials, the quadratic Gibbs kernel is equivalent to
$\exp(\langle x,y\rangle/\eps)$ (see, e.g., \cite{gushchin2024matching}). We use
\begin{equation}\label{eq:cond-kernel}
K_v^\eps(x,\dd y)
 =\frac{\exp(\langle x,y\rangle/\eps)v(y)\,\dd y}
 {Z_v^\eps(x)},\qquad
 Z_v^\eps(x)=\int_{\R^d}\exp(\langle x,y\rangle/\eps)v(y)\,\dd y.
\end{equation}
Writing the optimal plan of \eqref{eq:eot-primal} in Schr\"odinger form and
eliminating the potential associated with the \(\mu\)-marginal through its
marginal equation, the remaining potential \(v\) is characterized as the
minimizer of the semi-dual objective
\[
\mathcal J_{\mu,\nu}^\eps(v)
 =\int\log Z_v^\eps(x)\mu(\dd x)
 -\int\log v(y)\nu(\dd y),
\]
whose first-order optimality condition is precisely the \(\nu\)-marginal
constraint (see, e.g., \cite{nutz2022entropic}). Note that \(v\) is a
Schr\"odinger potential, not a probability density. In the implementation,
$v=v_\theta$ is a positive parametric function, such as a Gaussian mixture, with which the convergence of the optimization can be quite fast
\cite{korotin2024light}. 

\subsection{Notation and standing conventions}

Let \(\cP(\R^d)\) denote the probability measures on \(\R^d\), and for
\(q\geq1\) let
\[
 \cP_q(\R^d)
 :=\left\{\mu\in\cP(\R^d):
 \int_{\R^d}\|x\|^q\mu(\dd x)<\infty\right\}.
\]
For \(\mu,\nu\in\cP_2(\R^d)\), let \(\Pi(\mu,\nu)\) be the set of
couplings with marginals \(\mu\) and \(\nu\), and define
\[
 \Wtwo^2(\mu,\nu)
 :=\inf_{\pi\in\Pi(\mu,\nu)}
 \int_{\R^d\times\R^d}\|x-y\|^2\pi(\dd x,\dd y).
\]
For probability measures \(P\ll Q\), their Kullback--Leibler divergence is
\[
 \KL(P\Vert Q)
 :=\int\log\left(\frac{\dd P}{\dd Q}\right)\dd P.
\]
Moreover, their total
variation distance is
\[
 \|P-Q\|_{\mathrm{TV}}
 :=\sup_{A}|P(A)-Q(A)|.
\]
For
\(\sigma^2>0\) and \(z\in\R^d\), write
\[
 \phi_{\sigma^2}(z)
 :=(2\pi\sigma^2)^{-d/2}
 \exp\left(-\frac{\|z\|^2}{2\sigma^2}\right).
\]
For \(\mu\in\cP(\R^d)\), write
\[
 m_q(\mu):=\int_{\R^d}\|x\|^q\mu(\dd x),
\]
and set \(\log_+t:=\max\{\log t,0\}\). Write \(B_b(\R^d)\) for the bounded
measurable functions on \(\R^d\), and \(\mathrm{Lip}(\R^d)\) for the
globally Lipschitz functions, with \(\mathrm{Lip}(\psi)\) denoting the
Lipschitz constant of \(\psi\). The notation
\(a\lesssim b\) means \(a\leq Cb\) for a constant independent of the
level \(l\), and \(a\asymp b\) means \(a\lesssim b\) and \(b\lesssim a\).
Generic constants are denoted by \(C\); constants depending on the fixed
terminal time \(n\), the observation record, or the test function are
denoted by \(C_n\) and \(C_{n,\varphi}\), respectively.

\section{Method}\label{sec:method}

We first recall the bootstrap particle filter (PF) and its multilevel
variant (MLPF), and then introduce the sequential entropic optimal
transport filter (EOTF) together with its multilevel extension
(ML-EOTF).

\subsection{Particle filtering baselines}
\label{subsec:pf-baselines}

The bootstrap particle filter is the classical SMC approximation for
nonlinear, non-Gaussian state-space models\, see
Algorithm~\ref{alg:pf}. It approximates the prediction--Bayes recursion
by sequentially propagating particles, weighting them by the likelihood,
and resampling from the resulting weighted empirical distribution, and
may equivalently be viewed as an interacting-particle implementation of
the filtering Feynman--Kac flow
\cite{delmoral2004feynman}.

\begin{algorithm}[t]
\caption{Bootstrap particle filter (PF)}
\label{alg:pf}
\begin{algorithmic}[1]
\Require Initial particles \(\{x_0^{i,l}\}_{i=1}^N\), terminal time \(n\),
kernel \(M^l\), likelihoods \(\{G_p\}_{p=1}^n\)
\For{\(p=1,\ldots,n\)}
  \For{\(i=1,\ldots,N\)}
    \State Sample
    \(\widetilde x_p^{i,l}\sim M^l(x_{p-1}^{i,l},\cdot)\)
  \EndFor
  \State Set
    \(\displaystyle
    w_p^{i,l}
    =\frac{G_p(\widetilde x_p^{i,l})}
    {\sum_{j=1}^N G_p(\widetilde x_p^{j,l})}\),
    \(i=1,\ldots,N\)
  \State Draw ancestor indices
    \(\displaystyle
    a_p^{i,l}\sim\sum_{j=1}^N w_p^{j,l}\delta_j\),
    \(i=1,\ldots,N\)
  \For{\(i=1,\ldots,N\)}
    \State Set \(x_p^{i,l}=\widetilde x_p^{a_p^{i,l},l}\)
  \EndFor
\EndFor
\Ensure Terminal particles \(\{x_n^{i,l}\}_{i=1}^N\) and
\(\displaystyle
\filter_{n,\mathrm{PF}}^{l,N}(\varphi)
=\frac1N\sum_{i=1}^N\varphi(x_n^{i,l})\)
\end{algorithmic}
\end{algorithm}

The multilevel particle filter adapts the multilevel Monte Carlo idea
to SMC filtering \cite{jasra2017mlpf}. It combines a coarse-level PF
estimate with coupled fine/coarse level-difference estimators. For each
\(l\geq1\), the two particle systems are propagated through a coupled
kernel \(\check M^l\) with respective marginals \(M^l\) and \(M^{l-1}\),
such as the kernel induced by the Brownian coupling
\eqref{eq:brownian-coupling}. After likelihood weighting, the two sets
of ancestor indices are drawn from a coupling of the fine and coarse
categorical distributions, the maximal coupling being a common choice
(see Algorithm~\ref{alg:mlpf}). This preserves the two marginal
filtering approximations while reducing the variance of the level
difference.

\begin{algorithm}[t]
\caption{Multilevel particle filter (MLPF)}
\label{alg:mlpf}
\begin{algorithmic}[1]
\Require Finest level \(L\), terminal time \(n\), sample sizes
\(\{N_l\}_{l=0}^L\), kernels \(\{M^l\}_{l=0}^L\), coupled kernels
\(\{\check M^l\}_{l=1}^L\), likelihoods \(\{G_p\}_{p=1}^n\)
\State Run Algorithm~\ref{alg:pf} at level \(0\) with \(N_0\) particles
and set
\(\displaystyle
\widehat I_0^{\mathrm{PF}}
:=\filter_{n,\mathrm{PF}}^{0,N_0}(\varphi)\)
\For{\(l=1,\ldots,L\)}
  \For{\(i=1,\ldots,N_l\)}
    \State Draw \(x_0^{i,l}\sim\mu_0\) and set
    \(x_0^{i,l-1}=x_0^{i,l}\)
  \EndFor
  \For{\(p=1,\ldots,n\)}
    \For{\(i=1,\ldots,N_l\)}
      \State Sample
      \(\displaystyle
      (\widetilde x_p^{i,l},\widetilde x_p^{i,l-1})
      \sim\check M^l((x_{p-1}^{i,l},x_{p-1}^{i,l-1}),\cdot)\)
    \EndFor
    \State Set
      \(\displaystyle
      w_p^{i,l}
      =\frac{G_p(\widetilde x_p^{i,l})}
      {\sum_{j=1}^{N_l}G_p(\widetilde x_p^{j,l})}\),
      \(i=1,\ldots,N_l\)
    \State Set
      \(\displaystyle
      w_p^{i,l-1}
      =\frac{G_p(\widetilde x_p^{i,l-1})}
      {\sum_{j=1}^{N_l}G_p(\widetilde x_p^{j,l-1})}\),
      \(i=1,\ldots,N_l\)
    \State Draw ancestor-index pairs
    \(\{(a_p^{i,l},a_p^{i,l-1})\}_{i=1}^{N_l}\)
    from a maximal coupling of
    \(\{w_p^{j,l}\}_{j=1}^{N_l}\) and
    \(\{w_p^{j,l-1}\}_{j=1}^{N_l}\)
    \For{\(i=1,\ldots,N_l\)}
      \State Set
      \(x_p^{i,l}=\widetilde x_p^{a_p^{i,l},l}\)
      and
      \(x_p^{i,l-1}=\widetilde x_p^{a_p^{i,l-1},l-1}\)
    \EndFor
  \EndFor
  \State Set
    \(\displaystyle
    \widehat I_l^{\mathrm{PF}}
    :=\frac1{N_l}\sum_{i=1}^{N_l}
    \{\varphi(x_n^{i,l})-\varphi(x_n^{i,l-1})\}\)
\EndFor
\Ensure \(\displaystyle
\widehat\filter_{n,\mathrm{PF}}^{\mathrm{ML},L}(\varphi)
=\widehat I_0^{\mathrm{PF}}
+\sum_{l=1}^L\widehat I_l^{\mathrm{PF}}\)
\end{algorithmic}
\end{algorithm}

\subsection{Sequential EOT filter (EOTF)}

Given \(N\) particles \(\{x_0^{i,l}\}_{i=1}^N\) approximating the initial
filter \(\filter_0^l=\mu_0\), the EOTF uses the same propagation and
likelihood-weighting steps as the bootstrap PF in
Algorithm~\ref{alg:pf}. The difference lies in the update after
weighting: instead of drawing ancestor indices according to
\(\{w_p^{i,l}\}_{i=1}^N\), the EOTF constructs a transport from the
propagated particles to their likelihood-weighted counterpart. A fitted
Schr\"odinger potential \(v_{\theta_p^l}\) induces the conditional
kernel \(K_{v_{\theta_p^l}}^{\eps_l}\) (see definition in~\eqref{eq:cond-kernel}), and each propagated particle
\(\widetilde x_p^{i,l}\) generates one transported particle
\(x_p^{i,l}\). Likelihood information is thus incorporated by moving
the particles rather than duplicating highly weighted ones and
discarding the rest. The regularization \(\eps_l\) controls the
dispersion of the transport. The complete procedure is summarized in
Algorithm~\ref{alg:seq-eot}.

\begin{algorithm}[t]
\caption{Sequential EOT filter (EOTF)}
\label{alg:seq-eot}
\begin{algorithmic}[1]
\Require Initial particles \(\{x_0^{i,l}\}_{i=1}^N\), terminal time \(n\),
kernel \(M^l\), likelihoods \(\{G_p\}_{p=1}^n\), regularization
\(\eps_l\), potential family \(v_\theta\)
\For{\(p=1,\ldots,n\)}
  \For{\(i=1,\ldots,N\)}
    \State Sample
    \(\widetilde x_p^{i,l}\sim M^l(x_{p-1}^{i,l},\cdot)\)
  \EndFor
  \State Set
    \(\displaystyle
    w_p^{i,l}
    =\frac{G_p(\widetilde x_p^{i,l})}
    {\sum_{j=1}^N G_p(\widetilde x_p^{j,l})}\),
    \(i=1,\ldots,N\)
  \State Fit \(v_{\theta_p^l}\) by minimizing
    \(\displaystyle
    \mathcal J_{p,l}^N(\theta)
    =\frac1N\sum_{i=1}^N
    \log Z_{v_\theta}^{\eps_l}(\widetilde x_p^{i,l})
    -\sum_{i=1}^N
    w_p^{i,l}\log v_\theta(\widetilde x_p^{i,l})\)
  \For{\(i=1,\ldots,N\)}
    \State Sample
    \(\displaystyle
    x_p^{i,l}\sim
    K_{v_{\theta_p^l}}^{\eps_l}(\widetilde x_p^{i,l},\cdot)\)
  \EndFor
  \State Set
    \(\displaystyle
    \filter_p^{l,N}
    =\frac1N\sum_{i=1}^N\delta_{x_p^{i,l}}\)
\EndFor
\Ensure Terminal particles \(\{x_n^{i,l}\}_{i=1}^N\) and filtering estimate
\(\displaystyle
\filter_n^{l,N}(\varphi)
=\frac1N\sum_{i=1}^N\varphi(x_n^{i,l})\)
\end{algorithmic}
\end{algorithm}

\subsection{Scaled multilevel EOT coupling}
We now state how ML-EOTF works; see Algorithm~\ref{alg:ml-eot}.
For each fine/coarse pair set
\[
  \eps_l=\eps_0h_l^\kappa,
  \qquad \eps_0>0,\quad\kappa>0.
\]
Both levels start from the same initial particles. During prediction
they use the coupled transition in \eqref{eq:brownian-coupling}; each
marginal still uses its own kernel, weights, and fitted potential. At
the terminal time we separate potential fitting from particle
evaluation: the potential is fitted on a training ensemble and then
applied to an independent evaluation ensemble, with transport draws
independent across evaluation indices conditionally on the two
ensembles. This sample splitting is the mechanism used in
Section~\ref{sec:theory} to control cross-covariances; in practice it
can be replaced by the cross-fitting used in the experiments.

\begin{algorithm}[t]
\caption{Scaled multilevel EOT filter (ML-EOTF)}
\label{alg:ml-eot}
\begin{algorithmic}[1]
\Require Finest level \(L\), terminal time \(n\), sample sizes \(N_0\) and
\(\{(M_l,N_l)\}_{l=1}^L\), kernels \(\{M^l\}_{l=0}^L\), likelihoods
\(\{G_p\}_{p=1}^n\), regularizations \(\{\eps_l\}_{l=0}^L\), potential
family \(v_\theta\), observable \(\varphi\)
\State Run Algorithm~\ref{alg:seq-eot} at level \(0\) with \(N_0\)
particles and regularization \(\eps_0\); set
\(\displaystyle
\widehat I_0
:=\frac1{N_0}\sum_{i=1}^{N_0}\varphi(x_n^{i,0})\)
\For{\(l=1,\ldots,L\)}
  \State Independently run coupled copies of Algorithm~\ref{alg:seq-eot} at levels
  \(l-1\) and \(l\), using shared Brownian increments in
  \eqref{eq:brownian-coupling} and independent training/evaluation splits,
  to obtain training particles
  \(\{x_{n,\mathrm{tr}}^{j,l-1}\}_{j=1}^{M_l}\),
  \(\{x_{n,\mathrm{tr}}^{j,l}\}_{j=1}^{M_l}\) and evaluation particles
  \(\{x_{n,\mathrm{ev}}^{i,l-1}\}_{i=1}^{N_l}\)
  \State Fit \(\widehat v_{n,l}=v_{\theta_{n,l}}\) by minimizing
    \(\displaystyle \mathcal J_{n,l}^{\mathrm{pair},M_l}(\theta)
    =\frac1{M_l}\sum_{j=1}^{M_l}
    \log Z_{v_\theta}^{\eps_l}(x_{n,\mathrm{tr}}^{j,l-1})
    -\frac1{M_l}\sum_{j=1}^{M_l}
    \log v_\theta(x_{n,\mathrm{tr}}^{j,l})\)
  \For{\(i=1,\ldots,N_l\)}
    \State Sample
    \(\overline x_n^{i,l}\sim
    K_{\widehat v_{n,l}}^{\eps_l}(x_{n,\mathrm{ev}}^{i,l-1},\cdot)\)
  \EndFor
  \State Set
    \(\displaystyle
    \widehat I_l
    :=\frac1{N_l}\sum_{i=1}^{N_l}
    \{\varphi(\overline x_n^{i,l})
    -\varphi(x_{n,\mathrm{ev}}^{i,l-1})\}\)
\EndFor
\State \Return \(\displaystyle
\widehat\filter_n^{\mathrm{ML},L}(\varphi)
=\widehat I_0+\sum_{l=1}^L\widehat I_l\)
\end{algorithmic}
\end{algorithm}

The multilevel estimator is therefore
\[
 \widehat\filter_n^{\mathrm{ML},L}(\varphi)
 =\filter_n^{0,N_0}(\varphi)
 +\sum_{l=1}^L\frac1{N_l}\sum_{i=1}^{N_l}
 \{\varphi(\overline x_n^{i,l})-\varphi(x_{n,\mathrm{ev}}^{i,l-1})\}.
\]

\section{Theory}
\label{sec:theory}
In this section we mainly derive the theory for ML-EOTF; the single-level case (EOTF) follows from
similar---and strictly simpler---arguments.
We work conditionally on the fixed observation record \(y_{1:n}\). Recall
the Bayesian update map \(\Psi_p\) from \eqref{eq:bayes-map} and that
\[
 \filter_p^l=\Psi_p(\predictor_p^l),
 \qquad
 \predictor_p^l=\filter_{p-1}^lM^l.
\]
The terminal level difference is generated by coupling the limiting
marginal laws \(\filter_n^{l-1}\) and \(\filter_n^l\), with
\(\eps_l=\eps_0h_l^\kappa\) as in Algorithm~\ref{alg:ml-eot}.

The section is organized as follows. Section~\ref{sec:hypotheses} collects
the standing hypotheses. Section~\ref{sec:filter-discrepancy} derives the
fixed-time Wasserstein discrepancy between the fine and coarse filters.
Section~\ref{sec:eot-coupling} establishes the uniform density, entropy,
and moment regularity of the filtering laws and turns the \(O(\eps)\)
expansion of the quadratic EOT cost into a displacement bound.
Section~\ref{sec:particle-estimator} introduces the terminal sampling
scheme and proves the exact-potential variance bound under the
propagation-of-chaos condition \hypH{hyp:H-chaos}.
Section~\ref{sec:learned} extends the analysis to learned potentials, with
the learning error summarized by the joint Kullback--Leibler discrepancy
\(e_l\); a numerical rate for \(e_l\) is necessarily model-class and
optimizer dependent. Finally, Section~\ref{sec:complexity} upgrades the
one-step weak error to the filter level and states the main results of the article which is multilevel
complexity theorem (Theorem~\ref{thm:complexity}).
\subsection{Standing hypotheses}\label{sec:hypotheses}
We adopt the following conditions.
\begin{hypAitems}
  \item \label{hyp:A-dynamics}
  \textbf{State dynamics and moments.} The coefficients \(a,b\)
  are globally Lipschitz with linear growth,
  \(\int\|x\|^4\mu_0(\dd x)<\infty\), and the induced numerical filters
  satisfy
  \[
   \sup_{l\geq0}\sup_{p\leq n}
   \int\|x\|^4\filter_p^l(\dd x)\leq C_n.
  \]

  \item \label{hyp:A-likelihood}
  \textbf{Likelihood.} For every \(p\leq n\),
  \[
   0<\underline G_n\leq G_p(x)\leq\overline G_n<\infty,
   \qquad
   |G_p(x)-G_p(y)|\leq L_{G,n}\|x-y\|.
  \]

  \item \label{hyp:A-coupling}
  \textbf{Coupled strong error.} 
  There is $\beta>0$, such that
  for every \(x,y\), there is a
  coupled kernel \(\check M^l((x,y),\dd(x_f,x_c))\) with conditional
  marginals \(M^l(x,\cdot)\) and \(M^{l-1}(y,\cdot)\), satisfying
  \begin{equation}
   \int\|x_f-x_c\|^2\check M^l((x,y),\dd(x_f,x_c))
   \leq L_M^2\|x-y\|^2
   +C(1+\|x\|^2+\|y\|^2)h_l^{\beta}.
   \label{eq:coupled-kernel}
  \end{equation}

  \item \label{hyp:A-observable}
  \textbf{Observable.} The observable satisfies
  \(\varphi\in B_b(\R^d)\cap\mathrm{Lip}(\R^d)\), with Lipschitz constant
  \(L_\varphi\).

  \item \label{hyp:A-density}
  \textbf{Transition-density regularity.} For the density \(m^l(x,y)\) of the kernel
  \(M^l(x,\dd y)\), there are
  constants \(\texttt{C},\underline\sigma,\overline\sigma>0\),
  \(\texttt{L},\texttt{c}<\infty\), independent of \(l\), and a center map
  \(\mathfrak m_l:\R^d\to\R^d\) satisfying
  \(\|\mathfrak m_l(x)\|\leq \texttt{L}\|x\|+\texttt{c}\), such that
  \begin{equation}
   \texttt{C}^{-1}\phi_{\underline\sigma^2}(y-\mathfrak m_l(x))
   \leq m^l(x,y)
   \leq \texttt{C}~\phi_{\overline\sigma^2}(y-\mathfrak m_l(x)).
   \label{eq:kernel-gaussian-bound}
  \end{equation}

   \item \label{hyp:A-weakcost}
  \textbf{Weak error and cost.} There are
  \(\alpha,\gamma>0\) and \(s\geq0\) such that, for every globally
  Lipschitz test function \(\psi\),
  \[
   |M^l\psi(x)-M^\infty\psi(x)|
   \leq C_\psi(1+\|x\|^s)h_l^\alpha,
  \]
  and the total cost of one coupled level-\(l\) simulation, including
  potential fitting, satisfies \(C_l\lesssim h_l^{-\gamma}\).
\end{hypAitems}

The conditions are quite standard at fixed terminal
time in the particle filters literature, see \cite{jasra2017mlpf} and references therein.  For
  Euler--Maruyama under the Brownian coupling
  \eqref{eq:brownian-coupling}, one typically has \(\alpha=\beta=\gamma=1\) (see \cite{kloeden1992numerical}). One strengthening relative to a one-step estimate started at
\(x=y\) is \eqref{eq:coupled-kernel}, which allows arbitrary starting
points; this is needed because the fine and coarse filters no longer
coincide after the first update. Condition
\eqref{eq:kernel-gaussian-bound} is a fixed-time Aronson-type Gaussian
two-sided bound \cite{aronson1967bounds}. It is explicit for
linear-Gaussian dynamics and is available for standard nondegenerate
bounded-coefficient Euler--Maruyama schemes. The center map
\(\mathfrak m_l\) is not required to be the exact conditional mean; it is
the center of the Gaussian envelope.

\subsection{Filter discrepancy}\label{sec:filter-discrepancy}
In this section we bound the squared Wasserstein distance between the fine
and coarse filtering laws. We first check the following results.

\begin{lemma}
\label{prop:filter-stability}
Under \hypA{hyp:A-likelihood}, we have that
\[
 \Wtwo^2(\Psi_p(\mu),\Psi_p(\nu))
 \leq \left(\frac{\overline G_n}{\underline G_n}\right)^2
 \Wtwo^2(\mu,\nu),
 \qquad \mu,\nu\in\cP_2(\R^d).
\]
\end{lemma}

\begin{proof}
Let \(\pi\) be an optimal quadratic coupling of \((\mu,\nu)\). Define a
reweighted coupling by
\[
 \bar\pi(\dd x,\dd y)
 :=\frac{G_p(x)G_p(y)}{\mu(G_p)\nu(G_p)}\pi(\dd x,\dd y).
\]
Its two marginals are exactly \(\Psi_p(\mu)\) and \(\Psi_p(\nu)\), since
integrating in \(y\) gives \(G_p(x)\mu(\dd x)/\mu(G_p)\), and similarly for
the second marginal. Therefore
\[
\begin{aligned}
 \Wtwo^2(\Psi_p(\mu),\Psi_p(\nu))
 &\leq \int\|x-y\|^2\bar\pi(\dd x,\dd y)\\
 &\leq \frac{\overline G_n^2}{\underline G_n^2}
 \int\|x-y\|^2\pi(\dd x,\dd y),
\end{aligned}
\]
which proves the claim.
\end{proof}

\begin{lemma}
\label{prop:one-step}
Under \hypA{hyp:A-dynamics}--
\hypA{hyp:A-coupling}, there are
constants \(A_p,B_p<\infty\), independent of \(l\), such that
\[
 \Wtwo^2(\filter_p^l,\filter_p^{l-1})
 \leq A_p\Wtwo^2(\filter_{p-1}^l,\filter_{p-1}^{l-1})
 +B_ph_l^{\beta}.
\]
One may take
\(A_p=(\overline G_n/\underline G_n)^2L_M^2\).
\end{lemma}

The proof lifts the coupled kernel of \hypA{hyp:A-coupling} along an
optimal coupling of the previous filters and then applies
Lemma~\ref{prop:filter-stability}; the details are deferred to
Appendix~\ref{app:proofs}.

The following result is a straightforward consequence.
\begin{theorem}
\label{thm:filter-rate}
Under \hypA{hyp:A-dynamics}--
\hypA{hyp:A-coupling}, if
\(\filter_0^l=\filter_0^{l-1}=\mu_0\), then
\[
\Wtwo^2(\filter_n^l,\filter_n^{l-1})
 \leq C_nh_l^{\beta}.
\]
\end{theorem}

\begin{proof}
Iterate Lemma~\ref{prop:one-step}; the initial discrepancy vanishes. Precisely, we have that
\[
 \Wtwo^2(\filter_n^l,\filter_n^{l-1})
 \leq h_l^{\beta}
 \sum_{p=1}^nB_p\prod_{q=p+1}^nA_q.
\]
Setting $C_n:=\sum_{p=1}^nB_p\prod_{q=p+1}^nA_q$ proves the claim.
\end{proof}

\begin{remark}
The constant \(C_n=\sum_{p=1}^nB_p\prod_{q=p+1}^nA_q\) may grow geometrically
in \(n\), so all results in this paper are fixed-time statements. Long-time
or \(n\)-uniform estimates would additionally require stability of the
products \(\prod_qA_q\), for instance through filtering contraction or a
suitable forgetting property.
\end{remark}

\subsection{Exact-potential EOT coupling}\label{sec:eot-coupling}

The main external EOT input in this section is
\cite[Theorem~3.7]{malamut2025regularized}. For quadratic costs, and in
the form used here, that result gives an \(\mathcal O(\eps)\) expansion of the
transport-cost excess when the two marginals are absolutely continuous,
have finite absolute entropy, and possess moments of order
\(2+\delta\) for some \(\delta>0\). Proposition
\ref{prop:filter-regularity} verifies these three requirements uniformly
for the filtering laws, in the stronger
form of uniform absolute-entropy bounds. Theorem~\ref{thm:eot-displacement} then turns
that expansion into a displacement bound. The alternative logarithmic rate of \cite{eckstein2024quantization} is kept
only as a robustness remark and is not needed under
\hypA{hyp:A-density}.

Let \(\pi_{n,l}^{\eps_l}\) be the exact quadratic EOT plan with marginals
\((\filter_n^{l-1},\filter_n^l)\), and let
\[
 D_{n,l}^{\eps_l}(\varphi)
 =\varphi(X_f)-\varphi(X_c),
 \qquad
 (X_c,X_f)\sim\pi_{n,l}^{\eps_l}.
\]

\begin{proposition}[Uniform filtering regularity]
\label{prop:filter-regularity}
Under \hypA{hyp:A-dynamics}, \hypA{hyp:A-likelihood}, and
\hypA{hyp:A-density}, for every
\(1\leq p\leq n\) the filtering law \(\filter_p^l\) is absolutely
continuous, and
\[
 \sup_{1\leq p\leq n}\sup_{l\geq0}
 \left\{
 \int\|x\|^{2+\delta}\filter_p^l(\dd x)
 +\int \filter_p^l(x)|\log \filter_p^l(x)|\,\dd x
 \right\}\leq C_n
\]
for every \(0<\delta\leq2\).
\end{proposition}

The proof combines the Gaussian two-sided bound
\eqref{eq:kernel-gaussian-bound} with the uniform moment control of
\hypA{hyp:A-dynamics}; the details are deferred to
Appendix~\ref{app:proofs}.

We now consider the EOT plan \(\pi_{\mu,\nu}^{\eps}\) with
\[
 \mu=\filter_p^{l-1},
 \qquad
 \nu=\filter_p^l
\]
for simplicity. In this context, we claim the following result.
\begin{theorem}[Quadratic EOT displacement excess]
\label{thm:eot-displacement}
Under \hypA{hyp:A-dynamics}, \hypA{hyp:A-likelihood}, and
\hypA{hyp:A-density}, there exist
\(\eps_\star>0\) and \(C<\infty\) such that, for every
\(0<\eps\leq\eps_\star\), every \(p\leq n\), and every \(l\geq0\), we have
\[
 \int\|x-y\|^2\pi_{\mu,\nu}^{\eps}(\dd x,\dd y)
 \leq \Wtwo^2(\mu,\nu)+C\eps.
\]
\end{theorem}

\begin{proof}
By Proposition~\ref{prop:filter-regularity}, the marginals considered
here satisfy, uniformly in \(p\leq n\) and \(l\), the absolute-continuity,
absolute-entropy, and \(2+\delta\)-moment conditions required by
\cite[Theorem~3.7]{malamut2025regularized}. For
\(c(x,y)=\frac12\|x-y\|^2\), that theorem gives
\[
 \int c(x,y)\pi_{\mu,\nu}^{\eps}(\dd x,\dd y)
 \le \frac12\Wtwo^2(\mu,\nu)+C\eps,
\]
with a uniform constant. Multiplying by two proves the claim. 
\end{proof}

\begin{remark}
Without \hypA{hyp:A-density}, the quantization approach of
\cite{eckstein2024quantization} yields the weaker logarithmic bound
\[
 \int\|x-y\|^2\pi_{\mu,\nu}^{\eps}(\dd x,\dd y)
 \leq \Wtwo^2(\mu,\nu)+C\eps\log(1/\eps).
\]
All subsequent results extend to this regime with remainder
\(\mathcal O(\eps\log(1/\eps))\) in place of \(\mathcal O(\eps)\); we work under
\hypA{hyp:A-density} throughout. Expansions of the entropic transport
\emph{cost} itself, of a different nature, are obtained in
\cite{carlier2023rate}.
\end{remark}

\subsection{Particle-level exact-potential estimator}\label{sec:particle-estimator}

Let
\[
 Z_i=\varphi(\overline X_n^{i,l})-\varphi(X_n^{i,l-1}),
 \qquad
 \widehat D_{n,l}^{N_l,\eps_l}(\varphi)
 =\frac1{N_l}\sum_{i=1}^{N_l}Z_i.
\]
Since the first marginal of \(\pi_{n,l}^{\eps_l}\) is
\(\filter_n^{l-1}\), we may disintegrate it as
\[
 \pi_{n,l}^{\eps_l}(\dd x,\dd y)
 =
 \filter_n^{l-1}(\dd x)K_{n,l}^{\eps_l}(x,\dd y).
\]
We note that \(K_{n,l}^{\eps_l}\) is the conditional transport kernel from the
coarse terminal state to the fine terminal state.

For the exact-potential analysis, we isolate the terminal EOT pairing from
the finite-particle approximation of the preceding sequential filter. We
therefore consider the idealized terminal sampling scheme in which
\[
 X_n^{i,l-1}\sim\filter_n^{l-1},
 \qquad i=1,\ldots,N_l,
\]
marginally. The ensemble is generated by the interacting sequential EOT
filter of Algorithm~\ref{alg:seq-eot} and is therefore not independent
across \(i\); the residual cross-dependence is quantified by
\hypH{hyp:H-chaos} below. Conditionally on the terminal evaluation ensemble
\[
 \mathcal E_l:=\{X_n^{i,l-1}\}_{i=1}^{N_l},
\]
the transport draws are independent with
\[
 \overline X_n^{i,l}
 \sim K_{n,l}^{\eps_l}(X_n^{i,l-1},\cdot).
\]
Equivalently,
\[
 \law\left(
 \overline X_n^{1,l},\ldots,\overline X_n^{N_l,l}
 \,\middle|\,
 \mathcal E_l
 \right)
 =
 \prod_{i=1}^{N_l}
 K_{n,l}^{\eps_l}(X_n^{i,l-1},\cdot).
\]

We will use the following propagation-of-chaos condition on the evaluation
ensemble.

\begin{hypHitems}
  \item \label{hyp:H-chaos}
  There is a constant \(C_n\),
  independent of \(N_l\), \(h_l\), and \(\eps_l\), such that for every
  measurable function \(f:\R^d\to\R\) satisfying
  \(\filter_n^{l-1}(f^2)<\infty\),
  \[
   \left|
   \operatorname{Cov}\left(
   f(X_n^{i,l-1}),f(X_n^{j,l-1})
   \right)
   \right|
   \leq
   \frac{C_n}{N_l}\filter_n^{l-1}(f^2),
   \qquad i\neq j.
  \]
\end{hypHitems}

The exact filtering measures form a Feynman--Kac flow
\cite{delmoral2004feynman}, and propagation-of-chaos estimates of this type
are standard for selection--mutation Feynman--Kac particle systems
\cite{delmoral2018sharp}. 
Since the EOT update preserves the target marginal at the population level, the EOTF ensemble retains the mean-field interaction structure, with the empirical potential replacing the empirical weights; a dedicated propagation-of-chaos analysis is beyond the scope of this work. The $1/N_l$  variance scaling observed in Section~\ref{sec:numerics} is consistent with \hypH{hyp:H-chaos}.
Define
\begin{equation}\label{eq:F-def}
     F_{n,l}(x)
 :=
 \int_{\R^d}\{\varphi(y)-\varphi(x)\}K_{n,l}^{\eps_l}(x,\dd y),
 \qquad
 \xi_i:=Z_i-F_{n,l}(X_n^{i,l-1}).
\end{equation}
Under this condition, we obtain the
following estimates.

\begin{lemma}
\label{lem:chaos}
Under \hypH{hyp:H-chaos},
\[
 \left|\operatorname{Cov}(Z_i,Z_j)\right|
 \leq
 \frac{C_n}{N_l}\E|Z_1|^2,
 \qquad i\neq j.
\]
\end{lemma}

\begin{proof}
By construction,
\[
 Z_i=F_{n,l}(X_n^{i,l-1})+\xi_i,
 \qquad
 \E[\xi_i\mid\mathcal E_l]=0.
\]
The conditional independence of the transport draws gives
\[
 \E[\xi_i\xi_j\mid\mathcal E_l]=0,
 \qquad i\neq j.
\]
Since \(F_{n,l}(X_n^{i,l-1})\) is \(\mathcal E_l\)-measurable, all mixed
terms involving one \(F_{n,l}\) and one \(\xi\) also vanish after taking
conditional expectation. Hence
\begin{equation}
 \operatorname{Cov}(Z_i,Z_j)
 =
 \operatorname{Cov}\left(
 F_{n,l}(X_n^{i,l-1}),
 F_{n,l}(X_n^{j,l-1})
 \right).
 \label{eq:cov-reduction}
\end{equation}

Applying \hypH{hyp:H-chaos} to \(F_{n,l}\) gives
\[
 \left|\operatorname{Cov}(Z_i,Z_j)\right|
 \leq
 \frac{C_n}{N_l}\filter_n^{l-1}(F_{n,l}^2).
\]
Because \(X_n^{1,l-1}\sim\filter_n^{l-1}\), Jensen's inequality gives
\[
 \filter_n^{l-1}(F_{n,l}^2)
 =
 \E\left|F_{n,l}(X_n^{1,l-1})\right|^2
 \leq
 \E|Z_1|^2.
\]
This proves the pointwise covariance bound. 
\end{proof}

\begin{proposition}[Unbiasedness]
\label{prop:unbiased}
Under the terminal sampling scheme above,
\[
 \E\left[\widehat D_{n,l}^{N_l,\eps_l}(\varphi)\right]
 =
 \filter_n^l(\varphi)-\filter_n^{l-1}(\varphi).
\]
\end{proposition}

\begin{proof}
Since \(X_n^{i,l-1}\sim\filter_n^{l-1}\) and the second marginal of
\(\pi_{n,l}^{\eps_l}(\dd x,\dd y)
=\filter_n^{l-1}(\dd x)K_{n,l}^{\eps_l}(x,\dd y)\)
is \(\filter_n^l\), each \(Z_i\) has mean
\(\filter_n^l(\varphi)-\filter_n^{l-1}(\varphi)\).
\end{proof}

We now turn to second-moment estimates.
\begin{theorem}
\label{thm:exact-particle}
Under \hypA{hyp:A-dynamics}--
\hypA{hyp:A-density} and
\hypH{hyp:H-chaos},
\[
 \V\left[\widehat D_{n,l}^{N_l,\eps_l}(\varphi)\right]
 \leq
 \frac{C_{n,\varphi}}{N_l}
 \left(h_l^\beta+\eps_l\right).
\]

\end{theorem}

\begin{proof}
Fix \(i\). \hypA{hyp:A-observable} gives
\[
 |Z_i|^2
 \leq
 L_\varphi^2
 \left\|\overline X_n^{i,l}-X_n^{i,l-1}\right\|^2.
\]
Taking expectations and applying
Theorem~\ref{thm:eot-displacement} yields
\[
 \E|Z_i|^2
 \leq
 L_\varphi^2
 \left(
 \Wtwo^2(\filter_n^{l-1},\filter_n^l)+C\eps_l
 \right).
\]
By Theorem~\ref{thm:filter-rate}, 
\begin{equation}
 \V(Z_i)\leq\E|Z_i|^2
 \leq
 C_{n,\varphi}\left(h_l^\beta+\eps_l\right).
 \label{eq:single-particle-second-moment}
\end{equation}
Finally, expand the estimator variance as
\[
\V\left[\widehat D_{n,l}^{N_l,\eps_l}(\varphi)\right]
=
\frac1{N_l^2}\sum_{i=1}^{N_l}\V(Z_i)
+
\frac1{N_l^2}\sum_{i\neq j}\operatorname{Cov}(Z_i,Z_j).
\]
Using \eqref{eq:single-particle-second-moment} for the diagonal terms and
Lemma~\ref{lem:chaos} for the cross terms gives
\[
\begin{aligned}
\V\left[\widehat D_{n,l}^{N_l,\eps_l}(\varphi)\right]
&\leq
\frac{C_{n,\varphi}}{N_l}
 \left(h_l^\beta+\eps_l\right)
+
\frac{C_n}{N_l}\E|Z_1|^2\\
&\leq
\frac{C_{n,\varphi}}{N_l}
 \left(h_l^\beta+\eps_l\right),
\end{aligned}
\]
after enlarging \(C_{n,\varphi}\).
\end{proof}

\begin{corollary}
\label{cor:balance}
Under \hypA{hyp:A-dynamics}--
\hypA{hyp:A-density} and
\hypH{hyp:H-chaos}, with
\(\eps_l=\eps_0h_l^\kappa\).  Then
\[
 \V\left[\widehat D_{n,l}^{N_l,\eps_l}(\varphi)\right]
 \leq
 \frac{C_{n,\varphi,\eps_0}}{N_l}h_l^{\bar\beta},
 \qquad
 \bar\beta=\min\{\beta,\kappa\}.
\]
Thus the balanced choice is
\(
 \kappa=\beta.
\)

\end{corollary}

\begin{remark}[Non-adjacent pairs]
\label{rem:gap-pair}
The terminal EOT coupling is not restricted to adjacent levels. For
any \(l_f>l_c\), the same construction applied to the pair
\((\filter_n^{l_c},\filter_n^{l_f})\) at radius \(\eps_{l_f}\), combined
with the triangle inequality
\[
 \Wtwo(\filter_n^{l_c},\filter_n^{l_f})
 \leq\sum_{l=l_c+1}^{l_f}\Wtwo(\filter_n^{l-1},\filter_n^{l})
 \leq C_nh_{l_c}^{\beta/2},
\]
Theorem~\ref{thm:eot-displacement}, and the argument of
Theorem~\ref{thm:exact-particle}, yields
\[
 \V\left[\widehat D_{n,(l_f,l_c)}^{N,\eps_{l_f}}(\varphi)\right]
 \leq\frac{C_{n,\varphi}}{N}\left(h_{l_c}^{\beta}+\eps_{l_f}\right).
\]
A gap pair \((L,l_0)\) therefore carries the whole remaining
correction \(\filter_n^L(\varphi)-\filter_n^{l_0}(\varphi)\) at a
variance controlled by the coarse member \(l_0\) and essentially
independent of \(L\); this is the configuration used in
Section~\ref{sec:realdata}.
\end{remark}
\subsection{Learned-potential extension}\label{sec:learned}

We now allow the EOT potential to be learned from particles. Recall \(\pi_{n,l}^{\eps_l}\) is the
  exact EOT plan. Let
\(\widehat\pi_{n,l}^{\eps_l}\) denote the coupling induced by the learned
potential.  For $l\ge1$, define
  \begin{equation}\label{eq:el}
e_l:=\KL(\pi_{n,l}^{\eps_l}\Vert\widehat\pi_{n,l}^{\eps_l})
   +\KL(\widehat\pi_{n,l}^{\eps_l}\Vert\pi_{n,l}^{\eps_l}).
  \end{equation}
  For the level-$0$ estimator, writing $\widehat\filter_n^0$ for the
conditional law of a terminal EOTF particle given the training
ensemble $\mathcal T_0$, let
\[
e_0:=\KL(\filter_n^0\Vert\widehat\filter_n^0)
   +\KL(\widehat\filter_n^0\Vert\filter_n^0).
\]
Since the learned plans and $\widehat\filter_n^0$ depend on the training ensemble, each
\(e_l\), $l\ge0$, is random before conditioning on the training data. Unless
otherwise stated, expectations, variances, and mean squared errors in
Sections~\ref{sec:learned} and \ref{sec:complexity} are understood
conditionally on the collection of all training ensembles used to fit
the learned potentials. Under this conditioning, each learned plan and
hence each \(e_l\) is deterministic.
  Unlike the exact EOT coupling, \(\widehat\pi_{n,l}^{\eps_l}\) need not have exactly
\((\filter_n^{l-1},\filter_n^l)\) as its marginals. Gaussian mixtures can
nevertheless approximate the potential uniformly well in many scenarios
\cite{nguyen2020approximation,korotin2024light}, and quantitative stability
estimates control the plan-level error in terms of potential-level \cite{ghosal2022stability}.
Rather than pursuing a specific approximation class, we summarize all sources
of learning error by \(e_l\) and adopt the following moment condition.

\begin{hypLitems}
  \item \label{hyp:L-plan}
   For every realized learned plan under consideration,
  \[ 
   \int\|x-y\|^4\,
   \{\pi_{n,l}^{\eps_l}+\widehat\pi_{n,l}^{\eps_l}\}(\dd x,\dd y)
   \leq C_n.
  \]
\end{hypLitems}

We then obtain the following estimates.
\begin{lemma}
\label{lem:learned-kl}
Under \hypA{hyp:A-dynamics}, \hypA{hyp:A-likelihood},
\hypA{hyp:A-observable}, and \hypA{hyp:A-density}, together with
\hypL{hyp:L-plan},
\[
 |\E_{\widehat\pi_{n,l}^{\eps_l}}[\varphi(X_f)]
 -\filter_n^l(\varphi)|
 \leq C_{n,\varphi}e_l^{1/2},
 \qquad
 |\E_{\widehat\pi_{n,l}^{\eps_l}}[\varphi(X_c)]
 -\filter_n^{l-1}(\varphi)|
 \leq C_{n,\varphi}e_l^{1/2},
\]
and
\[
 \int\|x-y\|^2\widehat\pi_{n,l}^{\eps_l}(\dd x,\dd y)
 \leq \Wtwo^2(\filter_n^l,\filter_n^{l-1})
 +C\eps_l+C_ne_l^{1/4}.
\]

\end{lemma}

The proof combines Pinsker's inequality with a truncation of the
quadratic cost; the details are deferred to Appendix~\ref{app:proofs}.

\begin{remark}
The boundedness of \(\varphi\) in \hypA{hyp:A-observable} is used only
through Pinsker's inequality in the marginal bounds above. If \(\varphi\)
is merely globally Lipschitz, a truncation argument combined with the
second-moment control of \(\varphi\) under
\(\pi_{n,l}^{\eps_l}+\widehat\pi_{n,l}^{\eps_l}\) yields the same
conclusions with \(e_l^{1/4}\) in place of \(e_l^{1/2}\) in the marginal
bounds.
The same mechanism controls the level-$0$ learning error: 
\[
 \bigl|\E\!\left[\widehat I_0\,\middle|\,\mathcal T_0\right]
 -\filter_n^0(\varphi)\bigr|
 \leq C_{n,\varphi}\,e_0^{1/2}.
\]
\end{remark}

Combining the exact-potential analysis with
Lemma~\ref{lem:learned-kl} gives the following result.
\begin{theorem}[Learned-potential level estimator]
\label{thm:learned}
Under \hypA{hyp:A-dynamics}--
\hypA{hyp:A-density},
together with \hypH{hyp:H-chaos} and \hypL{hyp:L-plan},
\[
 \operatorname{MSE}[\widehat D_{n,l}^{N_l,\eps_l}(\varphi)]
 \leq Ce_l
 +\frac{C_{n,\varphi}}{N_l}
 \left(h_l^{\beta}+\eps_l+e_l^{1/4}\right).
\]
\end{theorem}

\begin{proof}
Write
\(\widehat Z_i=\varphi(\overline X_n^{i,l})-\varphi(X_{n,\mathrm{ev}}^{i,l-1})\)
for the summands produced by Algorithm~\ref{alg:ml-eot}; marginally,
\((X_{n,\mathrm{ev}}^{i,l-1},\overline X_n^{i,l})
\sim\widehat\pi_{n,l}^{\eps_l}\).
By the two marginal bounds of Lemma~\ref{lem:learned-kl},
\[
 \left|\E\bigl[\widehat D_{n,l}^{N_l,\eps_l}(\varphi)\bigr]
 -\bigl\{\filter_n^l(\varphi)-\filter_n^{l-1}(\varphi)\bigr\}\right|
 \leq C_{n,\varphi}e_l^{1/2},
\]
so the squared bias is at most \(Ce_l\). For the variance,
\hypA{hyp:A-observable} and the displacement bound of
Lemma~\ref{lem:learned-kl} give
\[
 \V(\widehat Z_i)
 \leq\E|\widehat Z_i|^2
 \leq L_\varphi^2\left\{
 \Wtwo^2(\filter_n^l,\filter_n^{l-1})+C\eps_l+C_ne_l^{1/4}\right\}
 \leq C_{n,\varphi}\bigl(h_l^{\beta}+\eps_l+e_l^{1/4}\bigr),
\]
using Theorem~\ref{thm:filter-rate} in the last step. Conditionally on the
training and evaluation ensembles, the learned transport draws are
independent across \(i\), so the covariance reduction
\eqref{eq:cov-reduction} applies verbatim with \(F_{n,l}\) in \eqref{eq:F-def} replaced by its
learned analogue, and \hypH{hyp:H-chaos} bounds each cross term by
\(\frac{C_n}{N_l}\E|\widehat Z_1|^2\) as in
Theorem~\ref{thm:exact-particle}. Summing the diagonal and cross terms and
adding the squared bias proves the claim.
\end{proof}

\subsection{Multilevel complexity}\label{sec:complexity}

It remains to combine the level-variance bounds with the weak-error and cost
rates of \hypA{hyp:A-weakcost}. We first upgrade the one-step weak error to
a fixed-time weak error for the filter itself, which closes the bias side of
the multilevel analysis. Write \(\filter_p^\infty\) for the exact filter
driven by \(M^\infty\).

\begin{proposition}[Fixed-time weak error]
\label{prop:weak-error}
Under \hypA{hyp:A-dynamics}, \hypA{hyp:A-likelihood}, and
\hypA{hyp:A-weakcost}, for every
\(\psi\in B_b(\R^d)\cap\mathrm{Lip}(\R^d)\),
\[
 |\filter_n^l(\psi)-\filter_n^\infty(\psi)|
 \leq C_{n,\psi}h_l^{\alpha}.
\]
\end{proposition}

The proof is a recursion over observation times combining the
bounded-Lipschitz stability of the Bayes update with the one-step weak
error of \hypA{hyp:A-weakcost}; the details are deferred to
Appendix~\ref{app:proofs}.

We can now state the complexity result.

\begin{theorem}[Multilevel complexity]
\label{thm:complexity}
Assume \hypA{hyp:A-dynamics}--\hypA{hyp:A-weakcost}, \hypH{hyp:H-chaos},
and \hypL{hyp:L-plan}. Let \(\eps_l=\eps_0h_l^\kappa\),
\(\bar\beta=\min\{\beta,\kappa\}\), and suppose the realized learning errors satisfy
\begin{equation}
 e_l^{1/4}\lesssim h_l^{\beta}+\eps_l,
 \qquad
 \sum_{l=0}^{L}e_l^{1/2}\lesssim h_L^{\alpha}.
 \label{eq:learning-control}
\end{equation}
For a target accuracy \(\delta>0\), choose the finest level \(L\) so that
\(h_L^{\alpha}\asymp\delta\), and allocate
\[
 N_l\propto\delta^{-2}\sqrt{V_l/C_l}\sum_{j=0}^{L}\sqrt{V_jC_j},
 \qquad
 V_l:=h_l^{\bar\beta},
 \qquad
 C_l\asymp h_l^{-\gamma},
\]
with \(\gamma\) as in \hypA{hyp:A-weakcost}. Then
\(\operatorname{MSE}\bigl[\widehat\filter_n^{\mathrm{ML},L}(\varphi)\bigr]
\lesssim\delta^2\),
where the error is measured against the exact filter
\(\filter_n^\infty\), at total cost
\[
 \operatorname{Cost}(\delta)=
 \begin{cases}
 \cO(\delta^{-2}),&\bar\beta>\gamma,\\
 \cO(\delta^{-2}|\log\delta|^2),&\bar\beta=\gamma,\\
 \cO(\delta^{-2-(\gamma-\bar\beta)/\alpha}),&\bar\beta<\gamma.
 \end{cases}
\]
\end{theorem}

\begin{proof}
Decompose the mean squared error against \(\filter_n^\infty(\varphi)\) into
variance and squared bias. For the variance, Theorem~\ref{thm:learned},
Corollary~\ref{cor:balance}, and the first condition in
\eqref{eq:learning-control} give
\(\V\bigl[\widehat D_{n,l}^{N_l,\eps_l}(\varphi)\bigr]\lesssim V_l/N_l\).
For the bias, the marginal bounds of Lemma~\ref{lem:learned-kl}
applied level by level, together with Proposition~\ref{prop:weak-error},
give
\[
 \bigl|\E\bigl[\widehat\filter_n^{\mathrm{ML},L}(\varphi)\bigr]
 -\filter_n^\infty(\varphi)\bigr|
 \lesssim\sum_{l=0}^{L}e_l^{1/2}+h_L^{\alpha}
 \lesssim\delta,
\]
using the second condition in \eqref{eq:learning-control} and the choice of
\(L\). The three cost regimes then follow from the standard multilevel
complexity theorem \cite{giles2015mlmc} with rate parameters
\((\alpha,\bar\beta,\gamma)\).
\end{proof}

For Euler--Maruyama, \(\alpha=\beta=\gamma=1\), so the balanced choice
\(\kappa=1\) places the estimator in the logarithmic complexity class
\(\cO(\delta^{-2}|\log\delta|^2)\): a near-\(\delta^{-2}\) cost for an
\(\cO(\delta)\) root-mean-square error, against the \(\cO(\delta^{-3})\)
cost of the single-level EOT filter.

\begin{remark}
Cost is measured in particle-steps, i.e.\ the number of propagated
particles times the number of time steps. The optimization overhead of the entropic
solver depends on adaptive iteration counts that vary with the level,
the regularization, and the particle budget, and is therefore hard to
quantify in closed form. We assess this empirically in Section~\ref{sec:numerics}.
\end{remark}

\section{Numerical results}
\label{sec:numerics}

We report three experiments. Section~\ref{sec:exp1} uses a
linear-Gaussian model of increasing dimension to illustrate the weight
degeneracy of the bootstrap particle filter and the benefit of replacing
resampling by a transport update. Section~\ref{sec:exp2} studies a
one-dimensional nonlinear diffusion on synthetic data, where the
potential optimization is cheap: this is the idealized regime of
Section~\ref{sec:theory}, and we verify the predicted variance decay and
complexity rates and demonstrate the necessity of the scaled schedule
$\eps_l=\eps_0 h_l^\kappa$. Section~\ref{sec:realdata} turns to a
five-asset stochastic-volatility model on real market data, where the
optimization overhead is no longer negligible and the theory-guided
hierarchy must be adapted: the production configuration couples a single
gap pair $(L,4)$, and the multilevel idea is shown to still improve the
accuracy--cost trade-off. Optimization
and implementation details are deferred to Appendix~\ref{app:implement}.

\subsection{PF versus EOTF in growing dimension}
\label{sec:exp1}

\paragraph{Setup.}
We consider a $d$-dimensional Ornstein--Uhlenbeck state-space model with
diagonal drift and full-rank observations,
\begin{equation}
  \mathrm{d}X_t = -a X_t\,\mathrm{d}t + \sigma\,\mathrm{d}W_t,
  \qquad
  Y_n = X_n + \eta_n, \quad \eta_n \sim \mathcal N(0, r^2 I_d),
  \label{eq:exp1-model}
\end{equation}
observed once per unit time. All coordinates are independent and
identically parameterized, so the exact Kalman filter is available per
coordinate and serves as ground truth. This model is a canonical stress
test for the bootstrap particle filter: the log-weight variance grows
linearly in $d$ ($\mathrm{Var}(\log w)\approx1.73\,d$ for the parameters
below), so the effective sample size collapses exponentially fast in the
dimension \cite{bengtsson2008curse}, and the particle budget would have to grow exponentially to
compensate.
We fix $a = 0.7$, $\sigma = 0.8$, $r = 0.45$, $X_0 \sim \mathcal
N(0, I_d)$, horizon $T = 50$, and $N = 150$ particles, and sweep
$d \in \{1, 2, 4, 8, 16, 32, 64\}$. The EOTF is run with $\varepsilon =
1.0$, $k = 3$ Gaussian-mixture components and $2$-fold cross-fitting of
the potential. All filters see the same observation sequence and results
are averaged over $50$ replications.

\paragraph{Metrics.}
We report (a) the RMSE of the filtering mean against the exact Kalman
mean; (b) the minimum ESS over time; (c) the minimum number of
\emph{distinct} particles (support size); and (d) the \emph{std ratio},
i.e.\ the terminal spread of the particle cloud (mean per-coordinate
standard deviation) relative to the exact Kalman posterior standard
deviation: a ratio near $1$ means calibrated uncertainty, $\ll 1$ a
collapsed, over-confident cloud.

\begin{table}[t]
  \centering
  \small
  \setlength{\tabcolsep}{4.5pt}
  \renewcommand{\arraystretch}{1.15}
  \begin{tabular}{|l|cccc|cccc|}
    \hline
    & \multicolumn{4}{c|}{bootstrap PF} & \multicolumn{4}{c|}{EOTF} \\
    \cline{2-9}
    $d$ & RMSE & min ESS & support & std ratio
        & RMSE & min ESS & support & std ratio \\
    \hline
    1  & \textbf{0.037} & 15.1 & 36.0 & 1.01 & 0.050 & 16.1 & 150 & 1.03 \\
    \hline
    2  & \textbf{0.077} & 3.2  & 14.6 & 0.99 & 0.083 & 3.1  & 150 & 1.02 \\
    \hline
    4  & \textbf{0.144} & 1.4  & 6.2  & 0.96 & 0.143 & 1.5  & 150 & 1.01 \\
    \hline
    8  & 0.263 & 1.1  & 2.7  & 0.95 & \textbf{0.244} & 1.1  & 150 & 1.01 \\
    \hline
    16 & 0.416 & 1.0  & 1.1  & 0.69 & \textbf{0.365} & 1.0  & 150 & 1.13 \\
    \hline
    32 & 0.553 & 1.0  & 1.0  & 0.39 & \textbf{0.476} & 1.0  & 150 & 1.24 \\
    \hline
    64 & 0.645 & 1.0  & 1.0  & 0.39 & \textbf{0.543} & 1.0  & 150 & 1.30 \\
    \hline
  \end{tabular}
  \caption{Linear-Gaussian results ($T=50$, $N=150$, mean over 50
  replications). EOTF with $\varepsilon=1.0$, $k=3$. Best RMSE in bold.}
  \label{tab:exp1}
\end{table}

\begin{figure}[t]
  \centering
  \IfFileExists{image/HD.pdf}
  {\includegraphics[width=0.95\textwidth,height=0.35\textwidth]{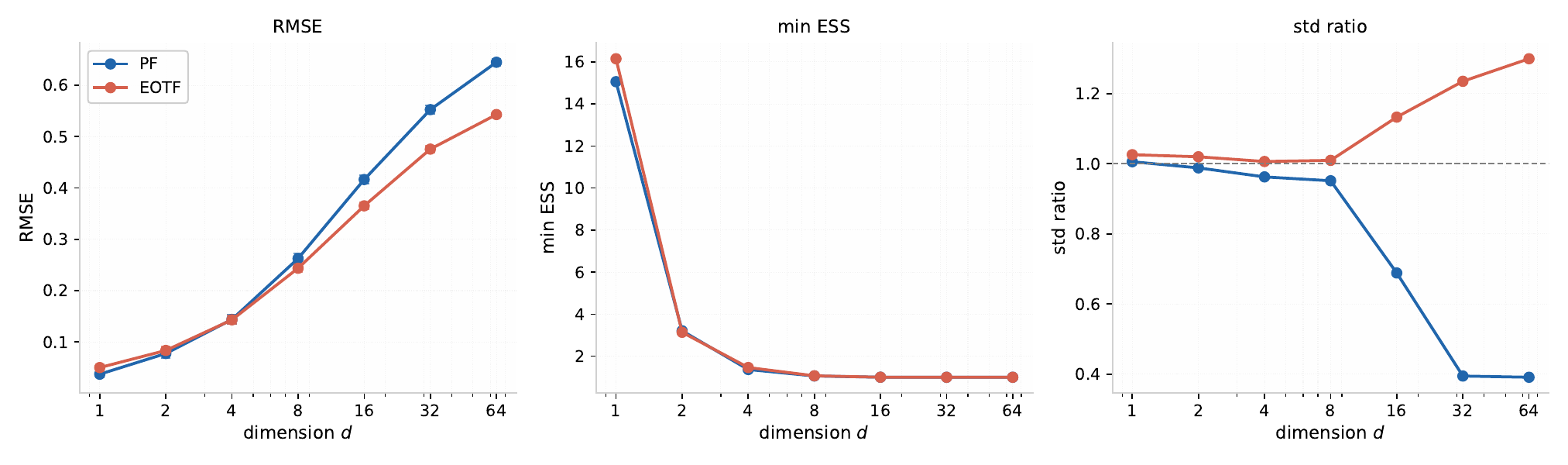}}
  {\fbox{\parbox[c][0.25\textheight][c]{0.9\textwidth}{\centering
   Placeholder for \texttt{image/HD.pdf}.}}}
  \caption{Linear-Gaussian model \eqref{eq:exp1-model} with $T=50$,
  $N=150$. Both filters suffer the same weight degeneracy
  ($\mathrm{ESS}\to 1$ for $d\ge 8$), but only the bootstrap PF's cloud
  disintegrates: its support shrinks to a single particle and its std
  ratio falls far below $1$, whereas EOTF retains all $150$ distinct
  transported particles with a std ratio close to $1$ throughout.}
  \label{fig:exp1}
\end{figure}

\paragraph{Results.}
Table~\ref{tab:exp1} and Figure~\ref{fig:exp1} show a crossover at
$d = 8$: the bootstrap PF is marginally better for $d \le 4$, where EOTF
only pays its small optimization-noise penalty, while for $d \ge 8$ EOTF
is consistently better, with a gap growing to $16\%$ at $d = 64$
($0.543$ vs $0.645$). The mechanism is visible in the ESS and support
columns, which separate two effects that are often conflated: the
concentration of the normalised weights (\emph{weight degeneracy}) and
its conversion, by resampling, into a collapse of the support
(\emph{sample impoverishment}). Once $\mathrm{Var}(\log w)\approx1.73d$
drives the weights of \emph{both} filters to $\mathrm{ESS}\approx1$,
neither method can extract further information from the current
particles --- the weight degeneracy is shared --- and only the response
differs. The bootstrap PF resamples from an essentially one-point weight
distribution, so its support disintegrates to $1$--$2$ distinct
particles and its error deteriorates beyond the no-information baseline.
EOTF never resamples: it fits the potential on the weighted empirical
measure and draws each new particle from the EOT conditional, so all
$150$ support points survive at every step and the error stays close to
the baseline. The std ratio sharpens the same conclusion: the PF cloud
is calibrated up to $d = 8$ and then collapses to $0.39\times$ the exact
posterior spread, while the EOTF cloud stays within $[1.0, 1.3]$ at
every dimension, only slightly over-dispersed due to the transport
noise.

\subsection{On one-dimensional synthetic diffusion}
\label{sec:exp2}

In dimension one the potential fits converge within a few dozen
iterations, so the entropic optimization is essentially free relative to
propagation and the experiment operates in the idealized regime of
Section~\ref{sec:theory}. We verify the variance decay of Corollary~\ref{cor:balance} and the complexity rates of Theorem~\ref{thm:complexity}. We use the following non-linear diffusion model (see \cite{jasra2017mlpf}).

\paragraph{Model.}
We consider a nonlinear mean-reverting diffusion with state-dependent
diffusion coefficient (NLM),
\begin{equation}
  \mathrm{d}X_t \;=\; \theta(\mu - X_t)\,\mathrm{d}t
                  \;+\; \frac{\sigma}{\sqrt{1 + X_t^2}}\,\mathrm{d}W_t,
  \qquad
  Y_n \mid X_{t_n} \;\sim\; \mathrm{Laplace}(X_{t_n},\, s),
  \quad t_n = n\,\Delta,\ n = 1,\dots,T,
  \label{eq:nlm}
\end{equation}
with parameters $\theta = 1$, $\mu = 0$, $\sigma = 0.5$, $s = 0.5$,
$\Delta = 1$ and horizon $T = 50$. The observable is $\varphi(x)=x$.
The factor $(1+X_t^2)^{-1/2}$ attenuates the diffusion at large $|X_t|$,
so the posterior is stable yet non-Gaussian; the model is thus a natural
stress test for the Gaussian-mixture potentials used by the EOT filter.
All replicates share the same synthetic observation path and differ only
in their simulation seeds.

\begin{figure}[t]
  \centering
  {\includegraphics[width=0.62\textwidth]{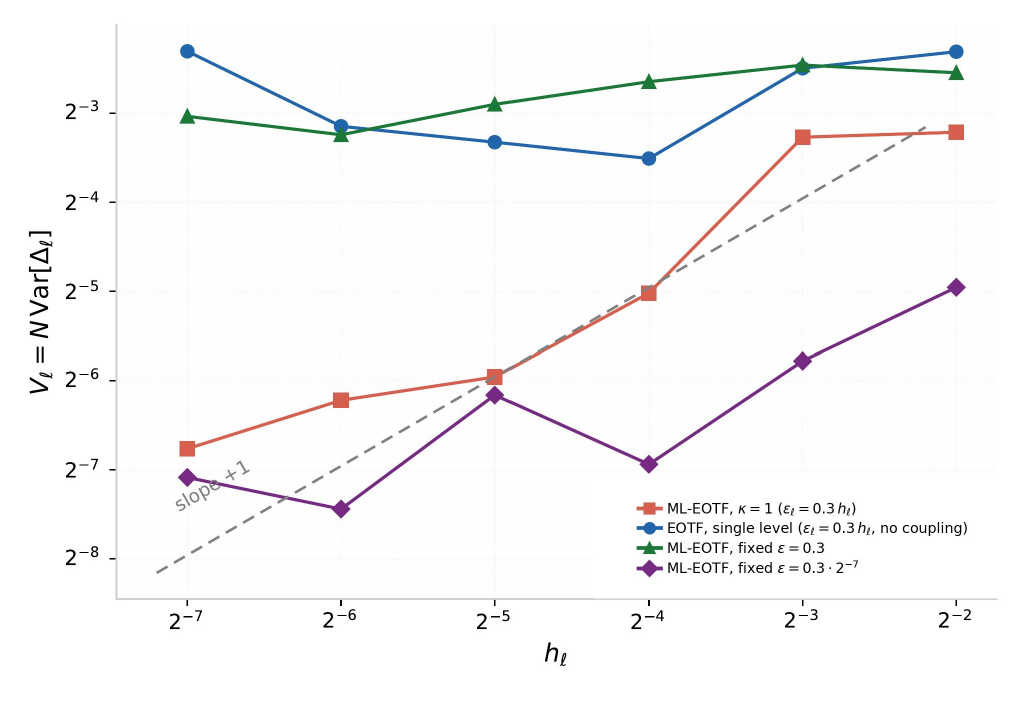}}

  \vspace{0.4cm}

  {\includegraphics[width=0.62\textwidth]{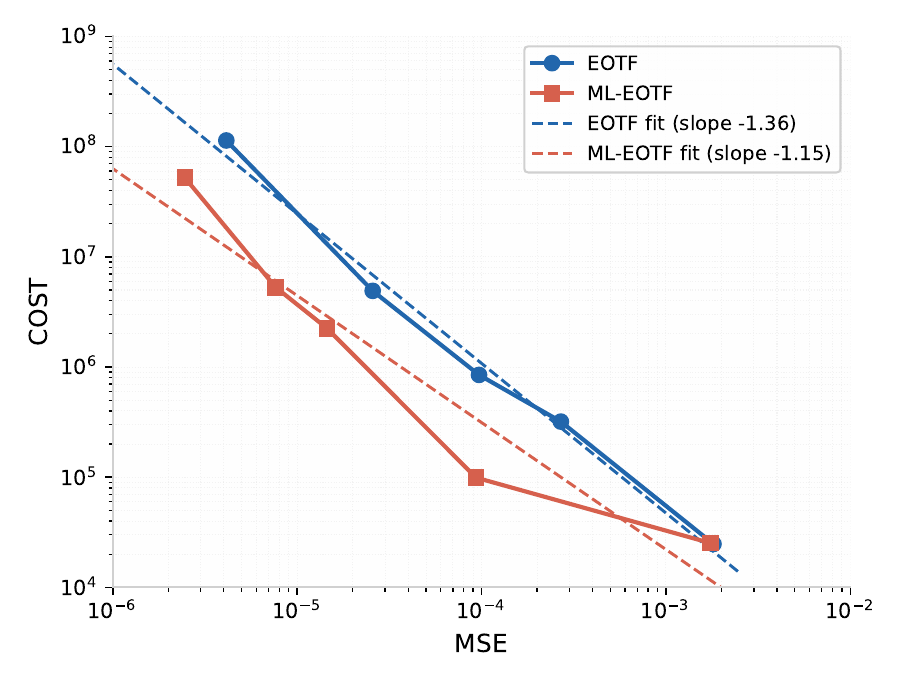}}

  \caption{Top: comparison of variance curves for the NLM model at
  $N = 500$, $\varepsilon_0 = 0.3$, $R = 50$ replicates: $\kappa = 1$
  schedule ($\varepsilon_l = 0.3\,h_l$), EOTF, ML-EOTF with fixed
  $\varepsilon = 0.3$, and ML-EOTF with fixed
  $\varepsilon = 0.3 \times 2^{-7}$; the dashed line has slope $+1$.
  Bottom: MSE against computational cost for the same model
  ($\varepsilon_0 = 0.3$, $l_{\max} \leq 7$): single-level EOTF and
  ML-EOTF frontiers with least-squares fits (dashed) of slopes $-1.36$
  and $-1.15$, respectively.}
  \label{fig:varcomp}
\end{figure}

\paragraph{Set-up.}
We fix $N = 500$ particles, $k = 3$ mixture components, and the
$\kappa = 1$ schedule $\varepsilon_l = \varepsilon_0 h_l$ with
$\varepsilon_0 = 0.3$, $h_l = 2^{-l}$, $l = 2,\dots,7$. Level
differences are computed with the coupled filter: shared Brownian
increments across levels, common random numbers in the transport
samplers, and a terminal coarse-to-fine EOT pairing with cross-fit
folds. All results use $R = 50$ independent replicates. The reference
solution is generated by a PF at level $l_{\max}=8$ with $N=2000$
particles. 
\paragraph{Results.}
\emph{Variance decay.}
Figure~\ref{fig:varcomp} (top) shows the measured $V_l$ against
$h_l$, where
\[
 V_l:=N\,\mathrm{Var}(\Delta_l),
 \qquad
 \Delta_l:=
 \frac1N\sum_{i=1}^N
 \{\varphi(\overline X_n^{i,l})
 -\varphi(X_{n,\mathrm{ev}}^{i,l-1})\}.
\]
For ML-EOTF with the $\kappa = 1$ schedule, the variance decays at a
rate close to $1$ ($\hat\beta = 0.98$ over $l = 3,\dots,6$ by
log--log least squares), in agreement with Corollary~\ref{cor:balance}.
The three alternatives at identical settings fail in distinct ways: the
single-level EOTF shows no visible decay of $V_l$ with $h_l$; the
multilevel filter with fixed $\varepsilon = 0.3$ exhibits a floor; and
with fixed $\varepsilon = \varepsilon_0 2^{-7}$ the decay is unstable.
This comparison demonstrates the necessity of scaling $\varepsilon_l$
with $h_l$.

\emph{MSE versus cost.}
Figure~\ref{fig:varcomp} (bottom) compares the MSE--cost frontiers of
EOTF and ML-EOTF, cost measured in particle-steps. The measured
behaviour is close to the rates predicted by
Theorem~\ref{thm:complexity} (frontier slopes $-1.36$ and $-1.15$ from
log--log fits, versus $-3/2$ and $-1$ up to the logarithmic
correction), and at any fixed accuracy ML-EOTF achieves a lower cost,
with savings of a factor $3$--$10$ for
$\mathrm{MSE} \lesssim 1.8\times10^{-3}$. 
Table~\ref{tab:wallclock} repeats the comparison in wall-clock
time, which shows
ML-EOTF is faster at every displayed accuracy, with a peak speedup of
$4.8\times$ around $\mathrm{MSE}\approx10^{-4}$. 

\begin{table}[t]
  \centering
  \caption{Wall-clock time (seconds, single platform) required by
  EOTF and ML-EOTF to reach a target MSE on the NLM model
  \eqref{eq:nlm} with the settings of Section~\ref{sec:exp2}.}
  \label{tab:wallclock}
  \small
  \setlength{\tabcolsep}{6pt}
  \renewcommand{\arraystretch}{1.15}
  \begin{tabular}{|c|cc|c|}
    \hline
    target MSE & EOTF (s) & ML-EOTF (s) & speedup \\
    \hline
    $1\times10^{-3}$ & 5.4   & 3.7  & $1.5\times$ \\
    \hline
    $5\times10^{-4}$ & 10.1  & 3.7  & $2.7\times$ \\
    \hline
    $2\times10^{-4}$ & 25.9  & 5.4  & $4.8\times$ \\
    \hline
    $1\times10^{-4}$ & 56.5  & 11.9 & $4.8\times$ \\
    \hline
    $5\times10^{-5}$ & 101.5 & 31.5 & $3.2\times$ \\
    \hline
  \end{tabular}
\end{table}

\subsection{On real data: a five-asset CIR stochastic-volatility model}
\label{sec:realdata}

\paragraph{Model and data.}
We consider a multi-asset Cox--Ingersoll--Ross (CIR) stochastic
volatility model on $d=5$ U.S.\ large-cap stocks (AAPL, AMZN, JPM, XOM,
JNJ). Each asset $i$ carries a variance factor $V^i_t$ following a CIR
process, which we express in normalised form $S^i_t = V^i_t/\theta_i$:
\begin{equation}\label{eq:cirsv}
  \mathrm{d}S^i_t = m_i\,(1 - S^i_t)\,\mathrm{d}t
  + \nu_i \sqrt{S^i_t}\,\mathrm{d}W^i_t,
  \qquad i=1,\dots,5,
\end{equation}
with independent Brownian motions $W_t^i$. The observation at day $n$ is
the vector of daily log-returns, modelled as
\begin{equation}\label{eq:obs}
  Y_n \;=\; D_n^{1/2}\, R^{1/2}\, \varepsilon_n \;+\; \tau\,\eta_n,
  \qquad
  D_n = \operatorname{diag}\big(\theta_i S^i_{t_n}\big),
\end{equation}
where $\varepsilon_n \sim \mathcal{N}(0, I_5)$ and
$\eta_n \sim \mathcal{N}(0, I_5)$ are independent,
$\tau = \operatorname{diag}(\tau_i)$, and $R$ is a correlation matrix.
The first term is the constant-conditional-correlation SV return: given
the volatility state, price innovations are correlated through $R$ while
the volatility factors evolve independently. The second term is an
idiosyncratic \emph{microstructure} component (bid--ask bounce,
non-synchronous trading) with scale tied to the long-run variance,
$\tau_i = c_\tau\sqrt{\theta_i}$, $c_\tau = 0.4$; such an additive
noise term is standard in realized-volatility and high-frequency SV
modelling. We use daily adjusted-close
log-returns, with an \emph{experiment window} of $T=49$ trading days
(2020-09-02 to 2020-11-10) disjoint from the data used to fit the
observation correlation and the initial state. The static parameters
$(m_i,\theta_i,\nu_i)$, are
estimated per asset by particle marginal Metropolis--Hastings; the
estimates are reported in Appendix~\ref{app:pmh}. They place all five
assets near or below the Feller boundary
($2m_i/\nu_i^2\in[0.81,\,1.08]$), so the standard strong order of
Euler-type schemes is unavailable.

\paragraph{Multilevel configuration.}
We discretise \eqref{eq:cirsv} on a mesh $h_l = 2^{-l}$ days by the
full-truncation Euler scheme (i.e., set $\sqrt{S^i_t}=0$ if $S^i_t<0$) and run the EOTF. The multilevel scheme couples levels $l-1$
and $l$ through shared Brownian increments and a terminal coarse-to-fine
EOT pairing at the fine-level entropic radius $\varepsilon_l=\varepsilon_0 h_l$ ($\kappa=1$,
$\varepsilon_0=0.3$).
Besides, we also consider a single \emph{gap pair}. Based on level $l=4$, we couple two levels $(L,4)$ with $L=5,6,7,8,9$, in which the
terminal coarse cloud of level $4$ is transported to the level-$L$
target at radius $\varepsilon_L$: the pair carries essentially the whole
bias correction at the price of one coarse-level fit (fine-level
information at a coarse-level price). This gap-pair construction is the non-adjacent variant of Algorithm~\ref{alg:ml-eot} analyzed in Remark~\ref{rem:gap-pair}. Each replicate is a complete filter run with
$N=500$ particles, so the multilevel sample sizes are replicate counts. The allocation is the cost-optimal choice given
the per-replicate variances and costs of Table~\ref{tab:real-levels}.
The reference value  is a level-$9$
fixed-$\varepsilon$ EOTF run ($N=1000$, $16$ replicates, Monte Carlo s.d.\
$1.3\times10^{-5}$); a scaled-$\varepsilon$ level-$9$ cross-check agrees
within Monte Carlo error.

\begin{figure}[t]
  \centering
  \includegraphics[width=0.49\textwidth]{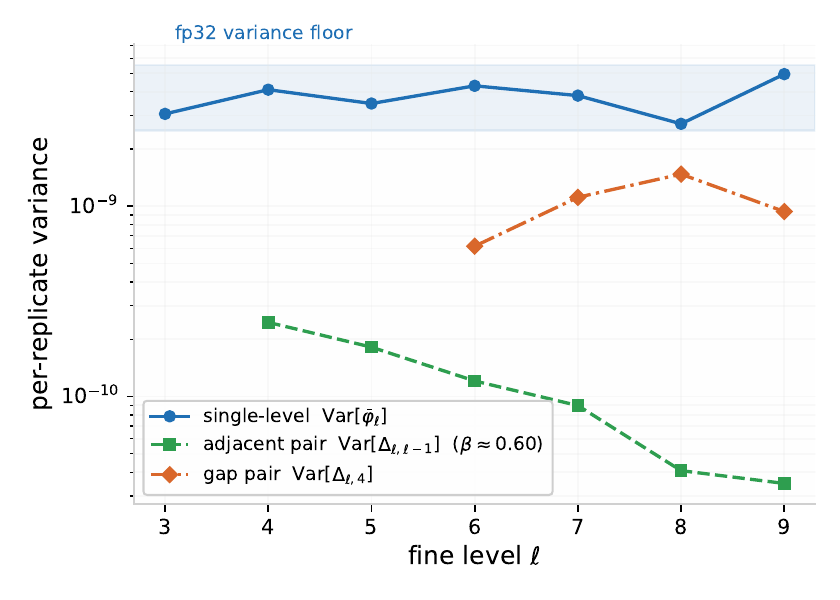}\hfill
  \includegraphics[width=0.49\textwidth]{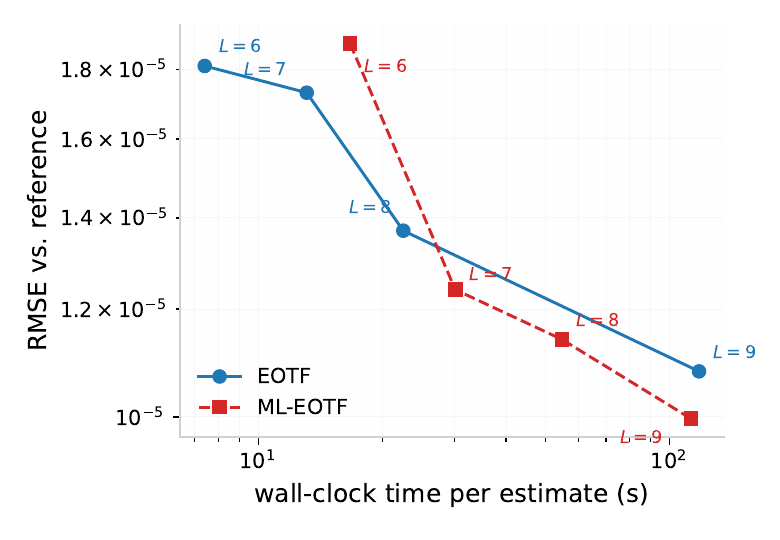}
  \caption{
  Left: per-replicate variances versus fine level $l$: single-level
  EOTF replicates, adjacent coupled pairs
  $(l,l-1)$ (decay rate $\beta\approx0.60$), and gap pairs
  $(l,4)$ used in production.
  Right: end-to-end RMSE versus wall-clock per estimate, $M=8$
  independent runs per marker, one marker per finest level
  $L\in\{6,7,8,9\}$.}
  \label{fig:real}
\end{figure}

\paragraph{Results.}
We repeated the full production experiment using wall-clock as cost.
First, the single-level per-replicate variance does \emph{not} decay
with $l$ as expected; the
coupling is unaffected, the adjacent-pair variance decaying with
measured rate $\beta \approx 0.60$; see Figure~\ref{fig:real} (left). However, according to Table~\ref{tab:real-levels}, the optimization takes a lot of computational time such that the rate of cost $\gamma$ is not equal to $1$ any more. Considering both variance and cost, we use \emph{gap pair} in ML-EOTF. The final result is shown in Table~\ref{tab:real-main}, which reports
the end-to-end accuracy--cost frontier, RMSE over $M=8$ independent
end-to-end estimates against the reference value (also see Figure~\ref{fig:real} (right)). At $L=6,7,8$ ML-EOTF estimator is
$2.3$--$2.4\times$ slower per estimate with comparable RMSE. As
$L$ grows the balance shifts: the cost of the multilevel estimator
grows more slowly than that of the level-$L$ EOTF run, and at $L=9$
ML-EOTF overtakes the single-level filter---RMSE
$9.97\times10^{-6}$ versus $1.08\times10^{-5}$ at $112.7$\,s versus
$117.9$\,s per estimate. We remark that if one can
speed up of the optimization stage---a faster optimizer or better GPU
utilization (e.g., float64 hardware with stronger throughput), one can amplify the
multilevel advantage.

\paragraph{Limitations.} We note two limitations of the present
study. First, the variance reduction across levels saturates: beyond
a certain level the RMSE--level curve plateaus, because the residual
error is dominated by a level-independent bias component rather than
by Monte Carlo variance, so that further refinement or replication
brings little improvement. Second, the multilevel speed-up presumes
that coarse levels are cheap; when the optimisation stage
(the daily potential fits) accounts for a large share of the
wall-clock cost, as in the present experiment, the cost ratio between
levels is compressed and the advantage of ML-EOTF over the
single-level filter is correspondingly modest.

\begin{table}[t]
  \centering
  \begin{tabular}{cc l ccc c}
    \toprule
    $L$ & method & config & RMSE & bias & std & time/est (s) \\
    \midrule
    6 & EOTF & $R=18$ & $1.81\times 10^{-5}$ & $-1.53\times 10^{-5}$ & $1.03\times 10^{-5}$ & 7.4 \\
      & ML-EOTF & $R_b=27,\,R_p=7$ & $1.88\times 10^{-5}$ & $-1.26\times 10^{-5}$ & $1.49\times 10^{-5}$ & 16.7 \\
    \midrule
    7 & EOTF & $R=32$ & $1.73\times 10^{-5}$ & $-1.53\times 10^{-5}$ & $8.62\times 10^{-6}$ & 13.1 \\
      & ML-EOTF & $R_b=58,\,R_p=17$ & $1.24\times 10^{-5}$ & $-1.40\times 10^{-6}$ & $1.32\times 10^{-5}$ & 30.1 \\
    \midrule
    8 & EOTF & $R=45$ & $1.37\times 10^{-5}$ & $-1.10\times 10^{-5}$ & $8.68\times 10^{-6}$ & 22.5 \\
      & ML-EOTF & $R_b=126,\,R_p=35$ & $1.16\times 10^{-5}$ & $-7.59\times 10^{-6}$ & $6.42\times 10^{-6}$ & 54.7 \\
    \midrule
    9 & EOTF & $R=164$ & $1.08\times 10^{-5}$ & $1.63\times 10^{-6}$ & $1.47\times 10^{-5}$ & 117.9 \\
      & ML-EOTF & $R_b=254,\,R_p=52$ & $9.97\times 10^{-6}$ & $3.72\times 10^{-6}$ & $5.00\times 10^{-6}$ & 112.7 \\
    \bottomrule
  \end{tabular}
  \caption{End-to-end accuracy and wall-clock cost of EOTF and ML-EOTF
  on the five-asset CIR--SV experiment. RMSE, bias and std are over $M=8$ independent end-to-end
  estimates against the reference; time is per
  estimate. ML-EOTF couples the base level $l=4$ with the gap pair
  $(L,4)$. $R$: replicates of the level-$L$ EOTF;
$R_b$, $R_p$: replicates of the base level $l=4$ and of the gap
pair $(L,4)$.}
  \label{tab:real-main}
\end{table}

\begin{table}[t]
  \centering
  \begin{tabular}{c ccc c ccc}
    \toprule
    \multicolumn{4}{c}{single-level} & & \multicolumn{3}{c}{coupled pair $(l_f,l_c)$} \\
    \cmidrule(r){1-4} \cmidrule(l){5-8}
    $l$ & Var/rep & mean & $t$/rep (s) & $(l_f,l_c)$ & Var/rep & mean & $t$/rep (s) \\
    \midrule
    3 & $3.06\times 10^{-9}$ & $6.74\times 10^{-4}$ & 0.24 & (4,3) & $2.45\times 10^{-10}$ & $5.79\times 10^{-6}$ & 0.51 \\
    4 & $4.10\times 10^{-9}$ & $6.93\times 10^{-4}$ & 0.24 & (5,4) & $1.82\times 10^{-10}$ & $4.23\times 10^{-6}$ & 0.53 \\
    5 & $3.47\times 10^{-9}$ & $6.82\times 10^{-4}$ & 0.26 & (6,5) & $1.21\times 10^{-10}$ & $1.97\times 10^{-6}$ & 0.56 \\
    6 & $4.30\times 10^{-9}$ & $6.90\times 10^{-4}$ & 0.29 & (7,6) & $8.96\times 10^{-11}$ & $4.89\times 10^{-7}$ & 0.62 \\
    7 & $3.82\times 10^{-9}$ & $6.80\times 10^{-4}$ & 0.35 & (8,7) & $4.07\times 10^{-11}$ & $1.41\times 10^{-6}$ & 0.75 \\
    8 & $2.71\times 10^{-9}$ & $6.87\times 10^{-4}$ & 0.47 & (9,8) & $3.49\times 10^{-11}$ & $-1.25\times 10^{-7}$ & 1.00 \\
    \addlinespace[2pt]
    9 & $4.95\times 10^{-9}$ & $6.86\times 10^{-4}$ & 0.71 & (6,4) & $6.16\times 10^{-10}$ & $1.34\times 10^{-5}$ & 0.54 \\
    & & & & (7,4) & $7.78\times 10^{-10}$ & $1.92\times 10^{-5}$ & 0.61 \\
    & & & & (8,4) & $9.44\times 10^{-10}$ & $1.75\times 10^{-5}$ & 0.73 \\
    & & & & (9,4) & $7.61\times 10^{-10}$ & $2.40\times 10^{-5}$ & 0.98 \\
    \bottomrule
  \end{tabular}
  \caption{Per-level measurements with 60
  replicates each. Left: single-level estimates. Right: coupled level differences
  $\Delta_{l_f,l_c}$, including the adjacent-pair and gap-pair $(l,4)$. The single-level variance shows no decay
  (floor $\approx4\times10^{-9}$), while the adjacent-pair
  variance decays with rate $\beta\approx0.60$ (see Figure~\ref{fig:real}).}
  \label{tab:real-levels}
\end{table}

\section{Conclusion}
\label{sec:conclusion}

We have introduced the sequential entropic optimal transport filter
(EOTF), which replaces the resampling step of the particle filter by
sampling from a learned EOT conditional kernel, and its multilevel
extension (ML-EOTF), in which fine and coarse terminal ensembles are
coupled through a scaled EOT plan with level-dependent regularization
\(\eps_l=\eps_0h_l^\kappa\). Under mild fixed-time conditions, the
level-difference estimator was shown to have variance of order
\(N_l^{-1}\{h_l^{\beta}+\eps_l\}\), with the learning error of the
fitted potentials summarized by a joint Kullback--Leibler discrepancy
whose contribution to the mean squared error is explicitly quantified.
For Euler--Maruyama discretizations this yields the canonical
multilevel complexity \(\cO(\delta^{-2}|\log\delta|^2)\) for an MSE of
\(\delta^2\). Numerically, the transport update prevents the support
collapse of the bootstrap particle filter in growing dimension, the
predicted variance and complexity rates are recovered on a nonlinear
diffusion, and on a five-asset stochastic-volatility problem the
multilevel construction, adapted through the gap-pair design of
Remark~\ref{rem:gap-pair}, still improves the accuracy--cost trade-off
once the finest level is reached.

Several directions remain open. First, our estimates are fixed-time
statements whose constants may grow geometrically in \(n\); one natural extension is the long-time
or \(n\)-uniform version of these estimates.
Second, the propagation-of-chaos condition \hypH{hyp:H-chaos} on the
EOTF ensemble has been assumed rather than proved; a dedicated
mean-field analysis of the transport-based update would complete the
particle-level theory. Third, quantitative rates for the learning
error \(e_l\) are necessarily model-class and optimizer dependent, and
deriving them for the Gaussian-mixture potentials used here would turn
conditions \eqref{eq:learning-control} into checkable prescriptions
for the training budgets. Finally, the experiments indicate
that the multilevel advantage is largest when coarse-level potential
fits are cheap; adaptive level schedules and non-adjacent pair designs
that allocate the optimization budget where it pays---as well as
localization ideas \cite{rebeschini2015can} for genuinely high-dimensional models---are natural
next steps.

\appendix
\section{Deferred proofs}
\label{app:proofs}

\subsection*{Proof of Lemma~\ref{prop:one-step}}

\begin{proof}
For \(\mu\in\cP_2(\R^d)\), recall
\[
 m_2(\mu)=\int\|x\|^2\mu(\dd x).
\]
Let \(\gamma_{p-1}\) be an optimal quadratic coupling between
\(\filter_{p-1}^l\) and \(\filter_{p-1}^{l-1}\), so that
\[
 \int\|x-y\|^2\gamma_{p-1}(\dd x,\dd y)
 =\Wtwo^2(\filter_{p-1}^l,\filter_{p-1}^{l-1}).
\]

For each pair of starting points \((x,y)\), let
\(\check M^l((x,y),\dd(x_f,x_c))\) be the coupled transition kernel from
\hypA{hyp:A-coupling}. Define a joint measure on the next
fine/coarse states by lifting this kernel along \(\gamma_{p-1}\):
\[
 \Lambda_p(\dd(x_f,x_c))
 :=\int_{\R^d\times\R^d}
 \check M^l((x,y),\dd(x_f,x_c))
 \,\gamma_{p-1}(\dd x,\dd y).
\]
We check the first marginal of \(\Lambda_p\) is
\[
 \int M^l(x,\dd x_f)\filter_{p-1}^l(\dd x)
 =\predictor_p^l(\dd x_f),
\]
and its second marginal is
\[
 \int M^{l-1}(y,\dd x_c)\filter_{p-1}^{l-1}(\dd y)
 =\predictor_p^{l-1}(\dd x_c).
\]
Therefore \(\Lambda_p\) is a coupling of the two predictors
\((\predictor_p^l,\predictor_p^{l-1})\). By the definition of the
Wasserstein distance,
\[
 \Wtwo^2(\predictor_p^l,\predictor_p^{l-1})
 \leq
 \int\|x_f-x_c\|^2\Lambda_p(\dd(x_f,x_c)).
\]
Using the definition of \(\Lambda_p\) and the pointwise coupled-kernel bound
\eqref{eq:coupled-kernel}, we obtain
\[
\begin{aligned}
 &\Wtwo^2(\predictor_p^l,\predictor_p^{l-1})\\
 &\quad\leq
 \int\left[
 L_M^2\|x-y\|^2
 +C(1+\|x\|^2+\|y\|^2)h_l^{\beta}
 \right]\gamma_{p-1}(\dd x,\dd y)\\
 &\quad=
 L_M^2\Wtwo^2(\filter_{p-1}^l,\filter_{p-1}^{l-1})
 +Ch_l^{\beta}
 \left\{
 1+m_2(\filter_{p-1}^l)+m_2(\filter_{p-1}^{l-1})
 \right\}.
\end{aligned}
\]
The uniform moment bound in \hypA{hyp:A-dynamics} therefore gives
\[
 \Wtwo^2(\predictor_p^l,\predictor_p^{l-1})
 \leq
 L_M^2\Wtwo^2(\filter_{p-1}^l,\filter_{p-1}^{l-1})
 +\widetilde B_ph_l^{\beta},
\]
where
\[
 \widetilde B_p
 :=
 C\left\{
 1+\sup_{j\geq0}m_2(\filter_{p-1}^j)
 +\sup_{j\geq0}m_2(\filter_{p-1}^{j-1})
 \right\}<\infty .
\]
Finally, since
\[
 \filter_p^l=\Psi_p(\predictor_p^l),
 \qquad
 \filter_p^{l-1}=\Psi_p(\predictor_p^{l-1}),
\]
Lemma~\ref{prop:filter-stability} implies
\[
\begin{aligned}
 \Wtwo^2(\filter_p^l,\filter_p^{l-1})
 &\leq
 \left(\frac{\overline G_n}{\underline G_n}\right)^2
 \Wtwo^2(\predictor_p^l,\predictor_p^{l-1})\\
 &\leq
 A_p\Wtwo^2(\filter_{p-1}^l,\filter_{p-1}^{l-1})
 +B_ph_l^{\beta},
\end{aligned}
\]
with
\[
 A_p=
 \left(\frac{\overline G_n}{\underline G_n}\right)^2L_M^2,
 \qquad
 B_p=
 \left(\frac{\overline G_n}{\underline G_n}\right)^2\widetilde B_p.
\]
This proves the one-step recursion.
\end{proof}

\subsection*{Proof of Proposition~\ref{prop:filter-regularity}}

\begin{proof}
Fix \(1\leq p\leq n\) and \(l\geq0\). First, by
\hypA{hyp:A-dynamics} and Jensen's inequality,
\[
 \int\|x\|^{2+\delta}\filter_p^l(\dd x)
 \leq
 \left(\int\|x\|^4\filter_p^l(\dd x)\right)^{(2+\delta)/4}
 \leq C_n^{(2+\delta)/4},
 \qquad 0<\delta\leq2.
\]
It remains to prove the density and entropy bounds.

By \hypA{hyp:A-density}, the prediction distribution
\(\predictor_p^l=\filter_{p-1}^lM^l\) has density
\[
 r_p^l(y)
 :=\int_{\R^d}m^l(x,y)\filter_{p-1}^l(\dd x).
\]

For the upper bound, \eqref{eq:kernel-gaussian-bound} gives
\[
 m^l(x,y)
 \leq \texttt{C}\phi_{\overline\sigma^2}(y-\mathfrak m_l(x))
 \leq \texttt{C}(2\pi\overline\sigma^2)^{-d/2}.
\]
Hence, with
\[
 \overline{\texttt{C}}:=\texttt{C}(2\pi\overline\sigma^2)^{-d/2},
\]
we obtain the uniform pointwise bound
\begin{equation}
 r_p^l(y)\leq \overline{\texttt{C}}.
 \label{eq:predictor-upper}
\end{equation}

For the lower bound, the Gaussian lower bound in
\eqref{eq:kernel-gaussian-bound} gives
\[
 r_p^l(y)
 \geq
 \texttt{C}^{-1}(2\pi\underline\sigma^2)^{-d/2}
 \int\exp\left(
 -\frac{\|y-\mathfrak m_l(x)\|^2}{2\underline\sigma^2}
 \right)\filter_{p-1}^l(\dd x).
\]
Since
\(\|\mathfrak m_l(x)\|\leq \texttt{L}\|x\|+\texttt{c}\),
\[
\begin{aligned}
 \|y-\mathfrak m_l(x)\|^2
 &\leq 2\|y\|^2+2\|\mathfrak m_l(x)\|^2\\
 &\leq 2\|y\|^2+4\texttt{L}^2\|x\|^2+4\texttt{c}^2.
\end{aligned}
\]
Consequently, there is a constant \(C_0>0\), independent of \(p\leq n\)
and \(l\), such that
\begin{equation}
 r_p^l(y)
 \geq
 C_0^{-1}e^{-C_0\|y\|^2}
 \int e^{-C_0\|x\|^2}\filter_{p-1}^l(\dd x).
 \label{eq:predictor-lower-pre}
\end{equation}

It remains to bound the last expectation from below. Let
\(X\sim\filter_{p-1}^l\) and set
\[
 \texttt{M}_{2,n}
 :=\sup_{j\geq0}\sup_{q\leq n}
 \int\|x\|^2\filter_q^j(\dd x)<\infty,
 \qquad
 \texttt{R}_{n}:=2(1+\texttt{M}_{2,n}).
\]
By Markov's inequality,
\[
 \mathbb P(\|X\|^2>\texttt{R}_{n})
 \leq\frac{\E\|X\|^2}{\texttt{R}_{n}}
 \leq\frac{\texttt{M}_{2,n}}{2(1+\texttt{M}_{2,n})}
 \leq\frac12.
\]
Therefore
\[
 \mathbb P(\|X\|^2\leq \texttt{R}_{n})\geq\frac12,
\]
and hence
\[
\begin{aligned}
 \E e^{-C_0\|X\|^2}
 &\geq
 \E\left[e^{-C_0\|X\|^2}
 \mathbf 1_{\{\|X\|^2\leq \texttt{R}_{n}\}}\right]\\
 &\geq e^{-C_0\texttt{R}_{n}}
 \mathbb P(\|X\|^2\leq \texttt{R}_{n})\\
 &\geq \frac12e^{-2C_0(1+\texttt{M}_{2,n})}.
\end{aligned}
\]
Substituting this estimate into \eqref{eq:predictor-lower-pre} gives
\begin{equation}
 r_p^l(y)\geq c_n^{-1}e^{-C_n\|y\|^2}
 \label{eq:predictor-lower}
\end{equation}
for constants \(c_n,C_n>0\) independent of \(p\leq n\) and \(l\).

The filtering density is
\[
 f_p^l(y)
 :=\frac{G_p(y)r_p^l(y)}{Z_p^l},
 \qquad
 Z_p^l:=\int G_p(z)r_p^l(z)\,\dd z.
\]
Because \(r_p^l\) is a probability density and
\(\underline G_n\leq G_p\leq\overline G_n\), we have
\[
 \underline G_n\leq Z_p^l\leq\overline G_n.
\]
Combining this with \eqref{eq:predictor-upper} and
\eqref{eq:predictor-lower} gives
\begin{equation}
 c_n'^{-1}e^{-C_n'\|y\|^2}
 \leq f_p^l(y)
 \leq \overline{\texttt{C}}~\frac{\overline{G}_n}{\underline{G}_n}.
 \label{eq:filter-two-sided}
\end{equation}
In particular, \(\filter_p^l\) is absolutely continuous with density
\(f_p^l\).

Finally, the upper bound in \eqref{eq:filter-two-sided} gives
\[
 \log f_p^l(y)\leq\log [\overline{\texttt{C}}~\frac{\overline{G}_n}{\underline{G}_n}],
\]
while the lower bound gives
\[
 \log f_p^l(y)
 \geq -\log c_n'-C_n'\|y\|^2.
\]
Therefore there is \(C_n''<\infty\) such that
\[
 |\log f_p^l(y)|
 \leq C_n''(1+\|y\|^2).
\]
Using the uniform second-moment bound,
\[
\begin{aligned}
 \int f_p^l(y)|\log f_p^l(y)|\,\dd y
 &\leq C_n''
 \int(1+\|y\|^2)f_p^l(y)\,\dd y\\
 &\leq C_n'''\,,
\end{aligned}
\]
uniformly in \(p\leq n\) and \(l\geq0\). This completes the proof.
\end{proof}

\subsection*{Proof of Lemma~\ref{lem:learned-kl}}

\begin{proof}
Write \(\pi=\pi_{n,l}^{\eps_l}\) and \(\widehat\pi=\widehat\pi_{n,l}^{\eps_l}\),
with the first coordinate interpreted as the coarse variable and the second as
the fine variable. By the definition of \(e_l\) and Pinsker's inequality,
\begin{equation}
 \|\pi-\widehat\pi\|_{\mathrm{TV}}
 \leq \sqrt{2\KL(\pi\Vert\widehat\pi)}
 \leq Ce_l^{1/2},
 \label{eq:tv-from-kl}
\end{equation}
and the same bound holds with the roles of \(\pi\) and \(\widehat\pi\)
reversed. 
For the marginal bias, since \(\varphi\in B_b(\R^d)\) by
\hypA{hyp:A-observable}, applying \eqref{eq:tv-from-kl} directly to
\(\varphi\) gives
\[
 \left|\int \varphi\,\dd\widehat\pi-\int\varphi\,\dd\pi\right|
 \leq 2\|\varphi\|_{\infty}\|\pi-\widehat \pi\|_{\mathrm{TV}}
 \leq C_{n,\varphi}e_l^{1/2}.
\]
Applying this to the second marginal gives the fine-level bound, and applying
it to the first marginal gives the coarse-level bound.

For the displacement, put \(q(x,y)=\|x-y\|^2\) and \(q_R=q\wedge R^2\). Since
\(q_R\leq R^2\), \eqref{eq:tv-from-kl} yields
\[
 \left|\int q_R\,\dd\widehat\pi-\int q_R\,\dd\pi\right|
 \leq R^2\|\widehat\pi-\pi\|_{\mathrm{TV}}
 \leq CR^2e_l^{1/2}.
\]
Moreover \(q-q_R=(q-R^2)_+\leq q\mathbf 1_{\{q>R^2\}}\), so by
Cauchy--Schwarz, Markov's inequality, and \hypL{hyp:L-plan},
\[
\begin{aligned}
 \int (q-q_R)\,\dd(\pi+\widehat\pi)
 &\leq
 \left(\int q^2\,\dd(\pi+\widehat\pi)\right)^{1/2}
 \left(\{\pi+\widehat\pi\}(q>R^2)\right)^{1/2}\\
 &\leq C R^{-2}.
\end{aligned}
\]
Thus
\[
 \left|\int q\,\dd\widehat\pi-\int q\,\dd\pi\right|
 \leq C\left(R^2e_l^{1/2}+R^{-2}\right).
\]
Choosing \(R^2=e_l^{-1/4}\) gives \(O(e_l^{1/4})\). Finally,
Theorem~\ref{thm:eot-displacement} applied to the exact plan gives
\[
 \int q\,\dd\pi
 \leq \Wtwo^2(\filter_n^l,\filter_n^{l-1})+C\eps_l,
\]
and combining the two displays proves the displacement estimate.
\end{proof}

\subsection*{Proof of Proposition~\ref{prop:weak-error}}

\begin{proof}
For \(\psi\in B_b(\R^d)\cap\mathrm{Lip}(\R^d)\), write
\(\|\psi\|_{\mathrm{BL}}:=\|\psi\|_\infty+\mathrm{Lip}(\psi)\), and set
\[
 D_p:=\sup_{\|\psi\|_{\mathrm{BL}}\leq1}
 \bigl|\filter_p^l(\psi)-\filter_p^\infty(\psi)\bigr|,
 \qquad p\leq n,
\]
so that \(D_0=0\) since \(\filter_0^l=\filter_0^\infty=\mu_0\). For any
\(\mu,\nu\in\cP(\R^d)\), since \(G_p\geq\underline G_n\), adding and
subtracting a cross term in \(\Psi_p(\mu)(\psi)=\mu(G_p\psi)/\mu(G_p)\)
gives
\begin{equation}
 |\Psi_p(\mu)(\psi)-\Psi_p(\nu)(\psi)|
 \leq \underline G_n^{-1}|\mu(G_p\psi)-\nu(G_p\psi)|
 +\underline G_n^{-1}\|\psi\|_\infty|\mu(G_p)-\nu(G_p)|.
 \label{eq:psi-bl-stability}
\end{equation}
Decompose
\[
 \filter_p^l(\psi)-\filter_p^\infty(\psi)
 =\underbrace{\Psi_p(\filter_{p-1}^lM^l)(\psi)
 -\Psi_p(\filter_{p-1}^\infty M^l)(\psi)}_{=:A_p}
 +\underbrace{\Psi_p(\filter_{p-1}^\infty M^l)(\psi)
 -\Psi_p(\filter_{p-1}^\infty M^\infty)(\psi)}_{=:B_p}.
\]
For \(A_p\), note that the synchronous coupling of the Euler--Maruyama
recursion with Lipschitz coefficients in \hypA{hyp:A-dynamics} gives
\(\|M^lf\|_{\mathrm{BL}}\leq C\|f\|_{\mathrm{BL}}\) uniformly in \(l\), and
that \(\|G_p\psi\|_{\mathrm{BL}}\leq C_n\|\psi\|_{\mathrm{BL}}\) by
\hypA{hyp:A-likelihood}. Hence \eqref{eq:psi-bl-stability} yields
\[
 |A_p|\leq C_nD_{p-1}.
\]
For \(B_p\), applying \eqref{eq:psi-bl-stability} with
\(\mu=\filter_{p-1}^\infty M^l\) and \(\nu=\filter_{p-1}^\infty M^\infty\),
and using \(|\nu(G_p\psi)|\leq\overline G_n\|\psi\|_\infty\), reduces the
bound to terms of the form
\(|\filter_{p-1}^\infty(M^lg)-\filter_{p-1}^\infty(M^\infty g)|\) with
\(g\in\{G_p\psi,G_p\}\). By \hypA{hyp:A-weakcost},
\[
 \bigl|\filter_{p-1}^\infty(M^lg)-\filter_{p-1}^\infty(M^\infty g)\bigr|
 \leq C_g\int\bigl(1+\|x\|^s\bigr)\filter_{p-1}^\infty(\dd x)\,h_l^{\alpha}
 \leq C_{n,g}h_l^{\alpha},
\]
where \(\sup_{p\leq n}m_s(\filter_p^\infty)\leq C_n\) follows, for
\(s\leq4\), from the fourth-moment hypothesis in \hypA{hyp:A-dynamics}
together with standard moment estimates for the exact diffusion and the
boundedness of \(G_p\); for \(s>4\) it is an additional mild moment
assumption. Iterating \(D_p\leq C_nD_{p-1}+C_n'h_l^{\alpha}\) from
\(D_0=0\) proves the claim.
\end{proof}

\section{Implementation details of the numerical experiments}
\label{app:implement}

\subsection{Potential fitting and transport sampling}
\label{app:optim}

\paragraph{Parameterization and closed-form conditional.}
The Schr\"odinger potential is parameterized as a diagonal Gaussian
mixture with $k=3$ components,
\[
 v_\theta(y)=\sum_{j=1}^k a_j\,\phi_{\Sigma_j}(y-m_j),
 \qquad a=\mathrm{softmax}(\alpha),
\]
with unconstrained logits $\alpha$, means $m_j$ and log-diagonal entries
of $\Sigma_j$, so positivity is automatic. With the exp-kernel both the
normalizer and the conditional transport kernel are available in closed
form:
\[
 Z_{v}^{\eps}(x)=\sum_{j=1}^k a_j
 \exp\!\Big(\frac{\langle x,m_j\rangle}{\eps}
 +\frac{\langle x,\Sigma_j x\rangle}{2\eps^{2}}\Big),
 \qquad
 K_{v}^{\eps}(x,\cdot)
 =\sum_{j=1}^k w_j(x)\,
 \mathcal N\!\big(m_j+\Sigma_j x/\eps,\ \Sigma_j\big),
\]
with $w_j(x)\propto a_j\exp(\langle x,m_j\rangle/\eps+\langle x,\Sigma_j
x\rangle/(2\eps^{2}))$. The objective is
minimised by Adam (learning rate $0.05$) on minibatches of the two
clouds.

\paragraph{Cross-fitting.}
Each daily update uses two folds: the potential is fitted on one half of
the weighted cloud and transports the other half, then the roles are
swapped, so no particle is transported by a potential that was fitted on
it. This decouples the optimization noise from the evaluation noise and
removes the associated self-fit bias; the terminal coarse-to-fine
pairing uses the same cross-fit construction.

\paragraph{Warm starting and iteration budgets.}
In production runs each level is warm-started from its own previous-day
potential, and a single daily fitting subsample is shared between the
two coupled levels, which keeps both fits in the same basin of the
non-convex objective at a fraction of the cost of independent cold fits.
The iteration budget is adaptive: a patience of $50$ iterations without
relative improvement $10^{-4}$ of the objective terminates the fit, with
a reduced budget ($35$--$42$) on coarser levels, where $\eps$ is larger
and the daily potential update smaller. Cold fits are capped at $800$
iterations.

\paragraph{Annealing at small $\eps$.}
For cold fits at small $\eps$ the optimisation is continued through a
geometric $\eps$-schedule ($0.3\to\eps_l$, ratio $1/2$) within the same
iteration budget. This is essential: at $\eps\lesssim10^{-2}$ the
$1/\eps^{2}$ curvature of the exp-kernel objective makes a cold Adam run
stall far from the optimum.

\paragraph{Implementation.}
All filters are implemented in PyTorch with the $R$ replicates and both
coupled levels batched as tensor operations; GPU runs use the Apple M5 MPS backend
in float32.

\subsection{Parameter estimation for the real-data experiment}
\label{app:pmh}

The static parameters $(m_i,\theta_i,\nu_i)$ of \eqref{eq:cirsv},
with $\xi_i=\nu_i\sqrt{\theta_i}$, are estimated per asset by particle
marginal Metropolis--Hastings on the 2020--2021 daily returns: the exact
CIR transition density (noncentral chi-square) is used inside a
bootstrap-filter likelihood estimator, with common random numbers across
the chain for the less well mixing components. The observation
correlation $R$ is estimated from pre-window daily returns with $10\%$
shrinkage towards the identity, and the initial states $S^i_0$ are
centred on the 21-day realised variance preceding the experiment window.
Table~\ref{tab:pmh} reports the estimates for the five assets of the
experiment.

\begin{table}[h]
  \centering
  \caption{PMMH estimates for the five-asset CIR--SV model (daily
  units). \texttt{acc} is the average acceptance rate of the chain and
  $s_0^i$ the initial state.}
  \label{tab:pmh}
  \small
  \begin{tabular}{l cccc c cc}
    \toprule
    ticker & $m_i$ & $10^{4}\theta_i$ & $10^{3}\xi_i$ & $\nu_i$
           & $2m_i/\nu_i^{2}$ & acc & $s_0^i$ \\
    \midrule
    AAPL & 0.137 & 2.74 & 8.98 & 0.543 & 0.93 & 0.18 & 1.76 \\
    AMZN & 0.079 & 2.24 & 5.79 & 0.387 & 1.05 & 0.43 & 1.18 \\
    JPM  & 0.125 & 2.85 & 9.34 & 0.553 & 0.81 & 0.36 & 0.97 \\
    XOM  & 0.080 & 2.22 & 5.72 & 0.384 & 1.08 & 0.12 & 1.36 \\
    JNJ  & 0.105 & 1.65 & 6.38 & 0.496 & 0.85 & 0.09 & 0.31 \\
    \bottomrule
  \end{tabular}
\end{table}


\begin{thebibliography}{99}

\bibitem{aljarrah2023optimal}
{\sc Al-Jarrah}, M., {\sc Hosseini}, B., \& {\sc Taghvaei}, A.~(2023).
Optimal transport particle filters.
In \emph{2023 62nd IEEE Conference on Decision and Control (CDC)},
pages 6798--6805. IEEE.

\bibitem{aronson1967bounds}
{\sc Aronson}, D. G.~(1967).
Bounds for the fundamental solution of a parabolic equation.
\emph{Bulletin of the American Mathematical Society},
73(6), 890--896.

\bibitem{bengtsson2008curse}
{\sc Bengtsson}, T., {\sc Bickel}, P., \& {\sc Li}, B.~(2008).
Curse-of-dimensionality revisited: Collapse of the particle filter in
very large scale systems.
In \emph{Probability and Statistics: Essays in Honor of David A.
Freedman}, Vol.~2, pages 316--335. Institute of Mathematical Statistics.

\bibitem{bertsimas1997introduction}
{\sc Bertsimas}, D. \& {\sc Tsitsiklis}, J. N.~(1997).
\emph{Introduction to Linear Optimization}.
Athena Scientific: Belmont, MA.

\bibitem{cappe2005inference}
{\sc Capp\'e}, O., {\sc Moulines}, E., \& {\sc Ryd\'en}, T.~(2005).
\emph{Inference in Hidden Markov Models}.
Springer: New York.

\bibitem{carlier2023rate}
{\sc Carlier}, G., {\sc Pegon}, P., \& {\sc Tamanini}, L.~(2023).
Convergence rate of general entropic optimal transport costs.
\emph{Calculus of Variations and Partial Differential Equations},
62(4), 116.

\bibitem{corenflos2021differentiable}
{\sc Corenflos}, A., {\sc Thornton}, J., {\sc Deligiannidis}, G., \&
{\sc Doucet}, A.~(2021).
Differentiable particle filtering via entropy-regularized optimal transport.
In \emph{Proceedings of the 38th International Conference on Machine
Learning}, pages 2100--2111.

\bibitem{cuturi2013sinkhorn}
{\sc Cuturi}, M.~(2013).
Sinkhorn distances: Lightspeed computation of optimal transport.
In \emph{Advances in Neural Information Processing Systems 26},
pages 2292--2300.

\bibitem{delmoral2004feynman}
{\sc Del Moral}, P.~(2004).
\emph{Feynman--Kac Formulae: Genealogical and Interacting Particle
Systems with Applications}.
Springer: New York.

\bibitem{delmoral2018sharp}
{\sc Del Moral}, P. \& {\sc Jasra}, A.~(2018).
A sharp first order analysis of Feynman--Kac particle models, Part I:
Propagation of chaos.
\emph{Stochastic Processes and their Applications},
128(1), 332--353.

\bibitem{delmoral2026bridging}
{\sc Del Moral}, P., {\sc Jasra}, A., \& {\sc Zhao}, K.~(2026).
On bridging mixture distributions.
arXiv:2608.13383.

\bibitem{delmoral2026sinkhorn}
{\sc Del Moral}, P. \& {\sc Jasra}, A.~(2026).
New trends in the stability of Sinkhorn semigroups.
\emph{Journal of Theoretical Probability},
39, Article 77.

\bibitem{doucet2001sequential}
{\sc Doucet}, A., {\sc de Freitas}, N., \& {\sc Gordon}, N., eds.~(2001).
\emph{Sequential Monte Carlo Methods in Practice}.
Springer: New York.

\bibitem{eckstein2024quantization}
{\sc Eckstein}, S. \& {\sc Nutz}, M.~(2024).
Convergence rates for regularized optimal transport via quantization.
\emph{Mathematics of Operations Research},
49(2), 1223--1240.

\bibitem{ghosal2022stability}
{\sc Ghosal}, P., {\sc Nutz}, M., \& {\sc Bernton}, E.~(2022).
Stability of entropic optimal transport and Schr\"odinger bridges.
\emph{Journal of Functional Analysis},
283(9), 109622.

\bibitem{giles2008mlmc}
{\sc Giles}, M. B.~(2008).
Multilevel Monte Carlo path simulation.
\emph{Operations Research},
56(3), 607--617.

\bibitem{giles2015mlmc}
{\sc Giles}, M. B.~(2015).
Multilevel Monte Carlo methods.
\emph{Acta Numerica},
24, 259--328.

\bibitem{gordon1993novel}
{\sc Gordon}, N. J., {\sc Salmond}, D. J., \& {\sc Smith}, A. F. M.~(1993).
Novel approach to nonlinear/non-Gaussian Bayesian state estimation.
\emph{IEE Proceedings F---Radar and Signal Processing},
140(2), 107--113.

\bibitem{gregory2016metpf}
{\sc Gregory}, A., {\sc Cotter}, C. J., \& {\sc Reich}, S.~(2016).
Multilevel ensemble transform particle filtering.
\emph{SIAM Journal on Scientific Computing},
38(3), A1317--A1338.

\bibitem{gregory2017seamless}
{\sc Gregory}, A. \& {\sc Cotter}, C. J.~(2017).
A seamless multilevel ensemble transform particle filter.
\emph{SIAM Journal on Scientific Computing},
39(6), A2684--A2701.

\bibitem{gushchin2024matching}
{\sc Gushchin}, N., {\sc Kholkin}, S., {\sc Burnaev}, E., \&
{\sc Korotin}, A.~(2024).
Light and optimal Schr\"odinger bridge matching.
In \emph{Proceedings of the 41st International Conference on Machine
Learning}.

\bibitem{jasra2017mlpf}
{\sc Jasra}, A., {\sc Kamatani}, K., {\sc Law}, K. J. H., \&
{\sc Zhou}, Y.~(2017).
Multilevel particle filters.
\emph{SIAM Journal on Numerical Analysis},
55(6), 3068--3096.

\bibitem{jasra2018mlpf}
{\sc Jasra}, A., {\sc Kamatani}, K., {\sc Osei}, P. P., \&
{\sc Zhou}, Y.~(2018).
Multilevel particle filters: Normalizing constant estimation.
\emph{Statistics and Computing},
28(1), 47--60.

\bibitem{ub_filt}
{\sc Jasra}, A., {\sc Law}, K. J. H., \& {\sc Yu}, F.~(2022).
Unbiased filtering of a class of partially observed diffusions.
\emph{Advances in Applied Probability},
54(3), 661--687.

\bibitem{kloeden1992numerical}
{\sc Kloeden}, P. E. \& {\sc Platen}, E.~(1992).
\emph{Numerical Solution of Stochastic Differential Equations}.
Springer: Berlin.

\bibitem{korotin2024light}
{\sc Korotin}, A., {\sc Gushchin}, N., \& {\sc Burnaev}, E.~(2024).
Light Schr\"odinger bridge.
In \emph{The Twelfth International Conference on Learning Representations}.

\bibitem{leonard2014survey}
{\sc L\'eonard}, C.~(2014).
A survey of the Schr\"odinger problem and some of its connections with
optimal transport.
\emph{Discrete and Continuous Dynamical Systems. Series A},
34(4), 1533--1574.


\bibitem{malamut2025regularized}
{\sc Malamut}, H. \& {\sc Sylvestre}, M.~(2025).
Convergence rates of the regularized optimal transport: Disentangling
suboptimality and entropy.
\emph{SIAM Journal on Mathematical Analysis},
57(3), 2533--2558.

\bibitem{nguyen2020approximation}
{\sc Nguyen}, T. T., {\sc Nguyen}, H. D., {\sc Chamroukhi}, F., \&
{\sc McLachlan}, G. J.~(2020).
Approximation by finite mixtures of continuous density functions that
vanish at infinity.
\emph{Cogent Mathematics \& Statistics},
7, 1750861.

\bibitem{nutz2022entropic}
{\sc Nutz}, M. \& {\sc Wiesel}, J.~(2022).
Entropic optimal transport: Convergence of potentials.
\emph{Probability Theory and Related Fields},
184, 401--424.

\bibitem{peyre2019computational}
{\sc Peyr\'e}, G. \& {\sc Cuturi}, M.~(2019).
Computational optimal transport: With applications to data science.
\emph{Foundations and Trends in Machine Learning},
11(5--6), 355--607.

\bibitem{reich2013nonparametric}
{\sc Reich}, S.~(2013).
A nonparametric ensemble transform method for Bayesian inference.
\emph{SIAM Journal on Scientific Computing},
35(4), A2013--A2024.

\bibitem{rebeschini2015can}
{\sc Rebeschini}, P. \& {\sc van Handel}, R.~(2015).
Can local particle filters beat the curse of dimensionality?
\emph{The Annals of Applied Probability},
25(5), 2809--2866.

\bibitem{sen2018coupling}
{\sc Sen}, D., {\sc Thiery}, A. H., \& {\sc Jasra}, A.~(2018).
On coupling particle filter trajectories.
\emph{Statistics and Computing},
28(2), 461--475.






\end{thebibliography}
\end{document}